\documentclass[aps,prb,reprint,floats,floatfix]{revtex4-1}
\usepackage{amsmath}
\usepackage{color}
\usepackage{amssymb}
\usepackage{graphicx}
\usepackage{bm}
\usepackage{array}
\usepackage{dcolumn}
\usepackage{bm}

\usepackage{xr}

\newcolumntype{d}[1]{D{.}{.}{#1} }

\DeclareMathAlphabet{\mathcalstd}{OMS}{cmsy}{m}{n}
\newcommand{\calP}{\mathcalstd P}

\newcommand{\jena}{Institut f\"ur Festk\"orpertheorie und -optik and European Theoretical Spectroscopy Facility,
	Friedrich-Schiller-Universit\"at Jena, Max-Wien-Platz 1, 07743 Jena, Germany}

\newcommand{\gmsn}{Grupo de Materiais Semicondutores e Nanotecnologia (GMSN), Instituto Tecnol\'{o}gico de Aron\'{a}utica (ITA), 12228-900 S\~{a}o Jos\'{e} dos Campos/SP, Brazil}

\begin{document}
	\title{Quantization of spin Hall conductivity in two-dimensional topological insulators versus symmetry and spin-orbit interaction}
	\author{Filipe Matusalem}\email{filipematus@gmail.com, gmsn@ita.br}\affiliation{\gmsn}
	\author{Lars Matthes}\affiliation{\jena}
	\author{J\"urgen Furthm\"uller}\affiliation{\jena}
	\author{Marcelo Marques}\affiliation{\gmsn}
	\author{Lara K. Teles}\affiliation{\gmsn}
	\author{Friedhelm Bechstedt}\affiliation{\jena}
	
	
	\begin{abstract}
		The third-rank tensor of the static spin Hall conductivity is investigated for two-dimensional (2D) topological insulators by electronic structure calculations. 
		Its seeming quantization is numerically demonstrated for highly symmetric systems independent of the gap size. 
		2D crystals with hexagonal and square Bravais lattice show similar effects, while true rectangular translational symmetry yields conductivity values much below the quantum $e^2/h$. 
		Field-induced lifting the inversion symmetry does not influence the quantum spin Hall state up to band inversion but the conductivity quantization. Weak symmetry-conserving biaxial but also uniaxial strain has a minor influence as long as inverted gaps dictate the topological character. The results are discussed in terms of the atomic geometry and the Rashba contribution to the spin-orbit interaction (SOI). Translational and point-group symmetry as well as SOI rule the deviation from the quantization of the spin Hall conductance.
	\end{abstract}
	
	\maketitle

\section{Introduction}

Two-dimensional (2D) quantum spin Hall (QSH) or topological insulators represent a class of quantum materials with an
insulating bulk but spin-polarized gapless edge states, which show linearly crossing bands and are protected by
time-reversal symmetry (TRS) \cite{Hasan2010,Yan2012,Franz2013}. 
The edge states are believed to be responsible for a quantized conductance in HgTe/HgCdTe \cite{Koenig2007} and InAs/GaSb \cite{Du2015a} quantum wells at ultralow temperatures. 
The quantum state of a 2D topological insulator without edges in question has been, however, not directly proven experimentally.
There are only theoretical predictions of the quantization of the spin Hall (SH) conductivity \cite{Hirsch1999}
in graphene \cite{Kane2005a} and related materials with hexagonal basal plane \cite{Matthes2016}. 
Usually the classification of 2D crystals as QSH or topological insulators (TIs)
\cite{Liu2011a,Xu2013,Si2014,Qian2014a,Ma2015,Zhang2016b,Ma2016b} is based on the computation of the topological invariant $Z_2=1$ and/or the justification of helical gapless edge states with opposite spin \cite{Fu2007}.

Several graphene-like staggered group-IV materials such as silicene \cite{Liu2011a}, germanene
\cite{Liu2011a}, stanene \cite{Xu2013}, and plumbene \cite{Yu2017} with a fundamental gap opened by spin-orbit interaction (SOI) \cite{Matthes2013} but also their chemically functionalized counterparts \cite{Xu2013,Si2014} crystallize with a hexagonal Bravais lattice.
Even, despite perturbations due to the interaction with a hexagonal substrate, their QSH phase may survive
\cite{Amlaki2016,Matusalem2016}. 
In contrast to transition metal dichalcogenides (TMDCs) with hexagonal lattice \cite{Ma2016b} also TMDCs with square \cite{Ma2015} or rectangular \cite{Qian2014a,Ma2016b,Tang2017} Bravais lattice have been predicted to be QSH insulators up to room temperature.

The characterization of the QSH phase by a quantized SH conductivity is not obvious. 
Since spin currents do not couple directly to experimental probe, it is difficult to measure them. Nevertheless, SH effects have been investigated experimentally \cite{Sinova2015}. 
The SH conductivity has been measured for several three-dimensional (3D) crystals, in particular metals \cite{Valenzuela2006,Hoffmann2013}. 
Theoretical studies in terms of model Hamiltonians, e.g. generalized Haldane model \cite{Haldane1988} to include TRS-invariant SOI \cite{Kane2005}, indicate that symmetry plays an important role. 
If mirror symmetry is broken, either by an electric field or by interaction with a substrate, the Rashba contribution to SOI influences the quantization \cite{Kane2005a}. 
Considering the spin accumulation for spin in normal direction, Kane and Mele \cite{Kane2005} concluded that the QSH phase is not generally characterized by a quantized SH conductivity, only for some mirror-symmetric 2D crystals.

For TIs crystallizing in honeycomb lattices, by means of \emph{first-principles} electronic-structure calculations including SOI and non-collinear spins \cite{Matthes2016}, it was recently demonstrated that the static SH conductivity is seemingly quantized in units of $e^2/h$, the conductance quantum, while it vanishes for trivial 2D insulators. 
Therefore, in this work we study the influence of point-group and translational symmetry on the static SH conductivity.
In detail the consequences of the interplay between crystallography and electronic topology is
investigated. 
2D TIs with hexagonal, square or rectangular Bravais lattices are used as examples. 
The inversion symmetry is broken by a vertical electric field. 
The character of the SH conductivity as a third-rank tensor and, hence, the spin orientation is taken into account. 
Possible electronic phase transitions are modeled by varying the Fermi energy or applying a biaxial or uniaxial strain. 
Germanene and stanene together with their derivatives functionalized by fluorine or iodine represent the hexagonal crystals. 
2D crystals with square and rectangular lattices are simulated by 1S and 1T' polymorphs of TMDCs, MoS$_2$ and WS$_2$. Model Hamiltonian studies are used to discuss the SOI influence on the static spin Hall conductivity. 

\section{Theoretical and Computational Methods}

We employ total-energy and electronic-structure calculations based on the density functional theory (DFT) as implemented in the Vienna ab-initio simulation package (VASP) \cite{Kresse1996b}. 
Exchange and correlation are described within the Perdew-Burke-Ernzerhof functional \cite{Perdew1996}.
Wave functions and pseudopotentials are generated within the projector-augmented wave method \cite{Kresse1999}.
Besides scalar-relativistic effects, also the SOI and non-collinear spins are taken into account.
The 2D crystals are simulated within the supercell approach with sufficiently thick vacuum layers to avoid artificial interactions between the periodic images.

The topological character of the electronic structure of the 2D crystals and the presence of the QSH phase are determined by calculating the $Z_2$ topological invariant \cite{Fu2007,Kane2005,Fu2006,Fu2007a}.
For inversion-symmetric systems it is derived from the parity of the occupied Bloch states at the time-reversal invariant momentum (TRIM) points of the Brillouin zone (BZ) \cite{Fu2007}. 
The computation for systems without inversion symmetry is performed using the method of Yu et al. \cite{Yu2011} as implemented in the VASP code \cite{Matthes2016}. 
The third-rank tensor of the intrinsic spin Hall conductivity $\sigma^i_{jk}(\omega)$ is calculated within the Kubo formalism \cite{Matthes2016a}. 
We focus on the static limit $\omega\rightarrow0$, $\sigma^i_{jk}(\omega\rightarrow0)=\sigma^i_{jk}$,

\begin{equation}\label{eq1}
\sigma^i_{jk}=\frac{e^2\hbar}{A}\sum_{\bf k}\sum_\nu f(\varepsilon_\nu({\bf k}))\Omega^i_{jk}(\nu,{\bf k})
\end{equation}
with the spin Berry curvature tensor of the occupied Bloch-Pauli state $|\nu,{\bf k}\rangle$ with energy $\varepsilon_\nu({\bf k})$ and occupation number $f(\varepsilon_\nu({\bf k}))$ \cite{Matthes2016a,Yao2005,Guo2008},

\begin{equation}\label{eq2}
\Omega^i_{jk}(\nu,{\bf k})=-2{\rm Im}\sum_{\nu'}{'}\frac{\langle\nu{\bf k}|\frac{1}{i}v^i_j|\nu'{\bf k}\rangle
	\langle\nu'{\bf k}|v_k|\nu{\bf k}\rangle}{\left[\varepsilon_\nu({\bf k})-\varepsilon_{\nu'}({\bf k})\right]^2},
\end{equation}
where the velocity operator related to the spin current $\frac{1}{i}v^i_j=\frac{1}{2m}(1-\delta_{ij})(-1)^\calP[p_k,\hat{\sigma}_i]_+$
($p_k$ - component of momentum operator, $\hat{\sigma_i}$ - component of Pauli matrix vector) and that related to the charge current $v_k=\frac{1}{m}p_k$ have been introduced. 
The Cartesian coordinates $\{i,j,k\}=\{x,y,z\}$ and their actual permutation $\calP$ are applied. 
Since in systems with SOI spin is not a conserved quantity, the spin Berry curvature \eqref{eq2} is not directly
related to its ordinary counterpart. 
The third-rank tensor \eqref{eq1} is due to a spin current along the $j$-th direction with a Cartesian spin component $i$, which is generated by an electric field in $k$-th direction. 
For the 2D systems we assume the spin current and the electric field direction in the $xy$-plane of the atomic sheet with $j=x$ and $k=y$. 
For hexagonal and square 2D crystals, the artificial 3D superlattice crystals are hexagonal/trigonal and tetragonal,
respectively. 
In these cases, non-zero elements only occur for three different Cartesian components $i$, $j$, $k$
\cite{Seemann2015}.
With mirror symmetry it holds $\sigma^z_{xy}=-\sigma^z_{yx}=-\sigma^{-z}_{xy}$. 
In the case of rectangular 2D crystals resulting in orthorhombic/monoclinic superlattices, besides $\sigma^z_{xy}$ the elements $\sigma^x_{xy}$ and $\sigma^y_{xy}$ of the third-rank tensor may occur.
However, because of \eqref{eq2} $\sigma^x_{xy}=0$ holds. 
The second in-plane spin component $\sigma^y_{xy}$ should be also negligibly small. 
The convergence of $\sigma^z_{xy}$ depends on the size of the fundamental energy gap and the density of the $\bf{k}$-point sampling of the BZ.
We apply self-adapting BZ integration schemes\cite{Zhang2009b}. Other approaches to compute intrinsic spin Hall conductivities use an interpolation scheme based on maximally localized Wannier functions\cite{Ryoo2019}.

\section{Numerical Results and Discussion}
\subsection{Hexagonal Topological Insulators}

\begin{figure}[h]
	\includegraphics[width=\linewidth]{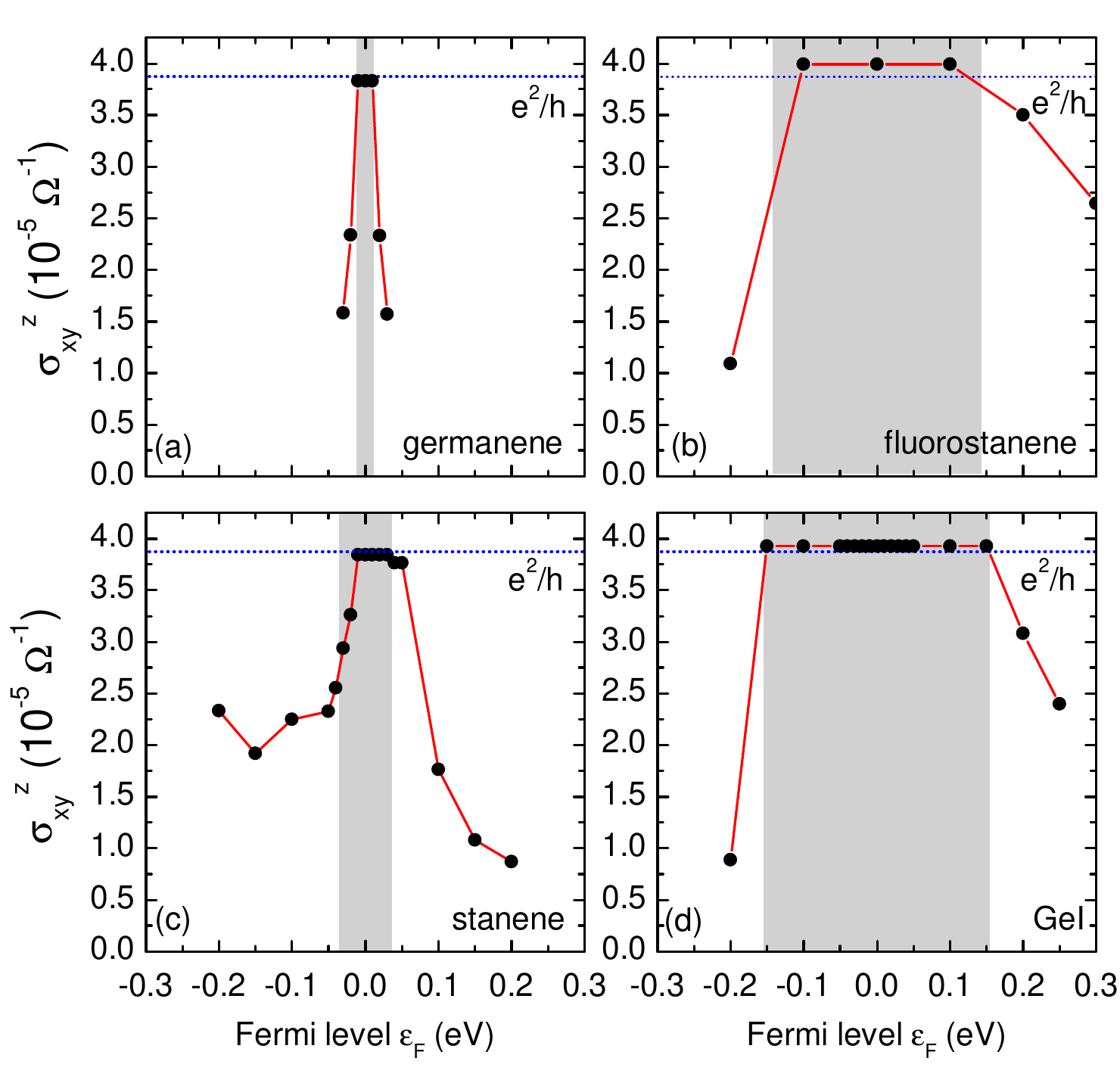} 
	\caption{Spin Hall conductivity versus Fermi level for (a) germanene, (b) fluorostanene, (c) stanene, and
		(d) GeI. The blue dotted line illustrates the quantum $e^2/h$. The hatched gray region indicates the band gap.
		Zero Fermi level is chosen in midgap position.}
	\label{fig1}
\end{figure}

We first investigate the static transverse SH conductivity $\sigma^z_{xy}$ \eqref{eq1} versus the position of the Fermi level for hexagonal crystals germanene, stanene, fluorostanene (SnF), and iodinated germanene (GeI) in Fig.~\ref{fig1}.

While the non-halogenated systems represent alternately buckled graphene-like sheets, halogen atoms are bonded to the group-IV basal plane in an alternating manner. 
Their symmetry and band gaps are indicated in Table~\ref{tab1}.
The 3D superlattice arrangements with the space group $D^3_{3d}$ $(P\bar{3}m1)$ and the point group $D_{3d}$ $(\bar{3}m)$ include inversion symmetry. 
Together with the inverted band structures the Fu-Kane parity method \cite{Fu2007} yields TIs with an invariant $Z_2=1$, despite the significant variation of the fundamental gap $E_g$ at $K$ or $\Gamma$ in the BZ between $E_g=24$ to 311~meV, in agreement with other studies \cite{Matthes2016,Xu2013,Si2014}.

\begin{table}[h!]
	\caption{Symmetry, band gap, and static spin Hall conductivity $\sigma^z_{xy}$ of TIs crystallizing in 2D
		hexagonal, square and rectangular Bravais lattices. The point group characterizes the 3D superlattice arrangement.
		The Fermi energy is in a midgap position.}
	\centering
	\begin{tabular}{|c|c|c|c|}
		\hline
		2D             & Point  & Band gap & $\sigma^z_{xy}$ \\
		system          & group  & (meV)    & $(10^{-5}\Omega^{-1})$ \\ \hline
		germanene       & $D_{3d}$ & 24.0  & 3.83 \\
		stanene         & $D_{3d}$ & 73.4  & 3.84 \\
		fluorostanene   & $D_{3d}$ & 287.7 & 3.99 \\
		GeI             & $D_{3d}$ & 310.6 & 3.93 \\
		1S-MoS$_2$      & $D_{4h}$ & 22.6  & 3.36 \\
		1S-WS$_2$       & $D_{4h}$ & 99.6  & 3.47 \\
		1T'-MoS$_2$     & $C_{2h}$ & 44.4  & -0.07 \\
		1T'-WS$_2$      & $C_{2h}$ & $\approx 0$  & 0.17 \\ \hline
	\end{tabular}
	\label{tab1}
\end{table}

Figure~\ref{fig1} underlines the topological character of pure and halogenated group-IV crystals as QSH insulators by the almost quantization $e^2/h$ of the SH conductivity $\sigma^z_{xy}$. 
When the Fermi energy $\varepsilon_F$ lies in the gap, the static conductivity is seemingly a quantized quantity with a value close to $\sigma^z_{xy}=e^2/h=3.874\times 10^{-5}\Omega^{-1}$, i.e., the reciprocal von Klitzing constant. The minor deviations of the computed $\sigma^z_{xy}$ are visible in Table \ref{tab1}. The computed values for germanene and stanene are smaller by less than 1 \% compared to $e^2/h$. The halogenated TIs GeI and SnF give rise to values slightly increased by 2 or 3 \%. This finding is maybe a consequence of the influence of the electron transfer from the basal plane to the F or I atoms on the SOI, in particular the Rashba interaction. For metallic systems with $\varepsilon_F$ in the conduction or valence bands, not only the insulating character but also the band topology is changed. The quantization of the SH conductivity is destroyed \cite{Matthes2016}. The midgap results of Fig. ~\ref{fig1} and Table~\ref{tab1}, however,  suggest the near conductivity quantization in hexagonal crystals, despite violation of general spin conservation by SOI and the absence of in-plane mirror symmetry, in contrast to model graphene\cite{Kane2005a}.

\subsection{Symmetry Reduction}

2D systems can be classified in four crystal systems, oblique, rectangular, square and hexagonal, five Bravais classes,
and 17 space groups, i.e., 17 crystal classes, which contain 10 different planar point groups \cite{bechstedt2012principles}.
Only 12 of the space groups may describe a topological system \cite{Dong2016,vanMiert2016,Hasan2010}.
Here, we study lower symmetries than the hexagonal graphene, which belongs to space group $D_{6h}$ $(P6/mmm)$, but include inversion.

Besides the first four hexagonal 2D TIs, Table~\ref{tab1} also shows results for 2D systems with square or rectangular Bravais lattice. 
Hexagonal crystals are represented in larger non-primitive unit cells with a rectangular lattice for test calculations. 
Despite different numerical procedures, e.g. {\bf k}-point samplings, the fundamental gaps and the SH conductivities
are conserved within the rectangular treatment.
This fact characterizes the quality of the self-adapting ${\bf k}$-point meshes.

2D TIs with a square Bravais lattice are rare. 
However, a new family of such topological systems has been predicted in monolayer TMDCs called 1S-MoS$_2$ and 1S-WS$_2$\cite{Ma2015,Nie2015,Sun2015}. 
They are three-layer stacks, wherein the transition metal atoms are sandwiched between sulphur atoms.
The four Mo or W and the eight S atoms per unit cell form the atomic basis of a square crystal with $D_{4h}$ point group (or $D_{2h}$ symmetry \cite{Sun2015}).
The tetragonal superlattice keeps this symmetry including the inversion.
The $Z_2$ invariants again provide direct evidence for the topological character. 
The two TIs have relatively sizeable band gaps with conduction band minima slightly out of the $\Gamma$ point with $E_g=23$~meV (MoS$_2$) or 99~meV (WS$_2$), which approach or exceed the thermal energy at room temperature. 

\begin{figure}[h]
	\includegraphics[width=\linewidth]{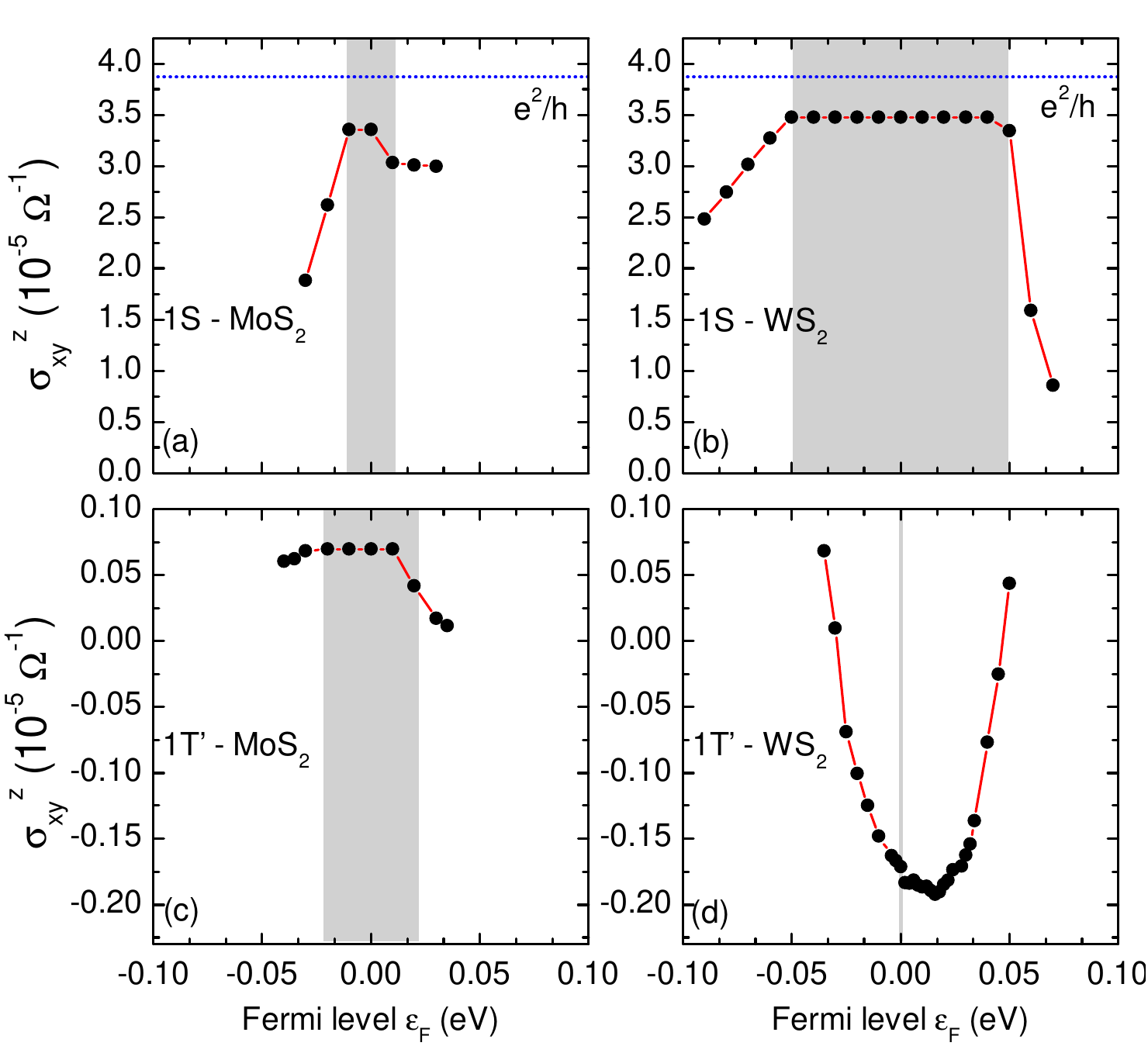} 
	\caption{Spin Hall conductivity versus Fermi level for (a) 1S-MoS$_2$, (b) 1S-WS$_2$, (c) 1T'-MoS$_2$ and (d) 1T'-WS$_2$. The blue dotted line illustrates the quantum $e^2/h$. The hatched gray region indicates the band gap.
		Zero Fermi level is chosen in midgap position. In the 1T' case the negative $\sigma^z_{xy}$ is plotted.}
	\label{fig2}
\end{figure}

The $\sigma^z_{xy}$ values in Table~\ref{tab1} and Figs.~\ref{fig2}a and b remain somewhat below the value $e^2/h$ of the reciprocal von~Klitzing constant. The small deviations of $\sigma^z_{xy}$ from the quantized value are a surprise because the normal spin component in $z$ direction is the only one, which should yield a nonzero conductivity for fixed planar Cartesian coordinates $x$ and $y$ due to the Hall geometry\cite{Seemann2015}. 
In a tetragonal system, the superlattice arrangement of the 2D square crystals, the components $\sigma^x_{xy}$ and $\sigma^y_{xy}$ of the third-rank tensor should vanish. 
Our findings $\sigma^z_{xy}\lesssim e^2/h$ in Figs.~\ref{fig2}a and b and Table~\ref{tab1}, i.e., no quantization, indicate that the quantized spin Hall conductivity is more sensitive to the interplay between crystallography and electronic topology \cite{Fu2007} than the topological invariant $Z_2=1$ computed from the parities at TRIM points $\Gamma$, $J$, $J$', and $K$ \cite{bechstedt2012principles} of the square BZ. 
In contrast to the Haldane model with up and down spins \cite{Haldane1988} and a quantized SH conductivity \cite{Kane2005a}, for the 1S-MoS$_2$ and 1S-WS$_2$ square lattices the normal component of the
spin is not conserved and the SH conductance $\sigma^z_{xy}$ is not quantized \cite{Kane2005}.
However, the in-plane rotational symmetry is smaller compared to the effective in-plane symmetry in hexagonal TIs. 

The polymorphism of monolayer TMDCs \cite{Qian2014a} allows to study also rectangular 2D crystals.
Indeed, the 1T' phases of the monolayer MX$_2$ with M=Mo, W and X=S, Se, Te are theoretically predicted to
be a promising new class of QSH insulators with large band gap. 
The three atomic planes X-M-X with rhombohedral stacking are unstable and undergo a spontaneous lattice distortion resulting in a period doubling in $x$ direction and zig-zag chains along the $y$ direction. 
The rectangular unit cells contain two M and four X atoms. We investigate 1T'-MoS$_2$ and 1T'-WS$_2$ as prototypical examples in Table~\ref{tab1} and Figs.~\ref{fig2}c and d. 
The atomic relaxation does not destroy inversion symmetry but leads to the $C_{2h}$ point group of the 3D arrangements. 1T'-MoS$_2$ has a  fundamental direct gap of 44~meV near to $\Gamma$ point at the $\Gamma$-$Y$  line, while 1T'-WS$_2$ has an almost vanishing  fundamental indirect gap between $\Gamma$ and a conduction band minimum position somewhat away from  the high-symmetry $\Gamma$-$Y$ line. The direct gaps at $\Gamma$ are 539 and 192~meV, respectively, close to results of other DFT calculations\cite{Qian2014a}. 
The two gaps characterize conduction and valence bands around $\Gamma$, with a camelback shape due to the $pd$ band inversion. 
Consequently, we confirm $Z_2=1$ and, therefore, the TI/QSH phase character. There are other rectangular 2D crystals, the group-IV monochalcogenides. They are trivial insulators with a large but predicted to exhibit a giant spin Hall effect\cite{Sawinska2019}. 

Figures~\ref{fig2}c and d and the small static SH conductivity values in Table~\ref{tab1} indicate drastic changes with respect to all other 2D crystals discussed above. 
Still, the value of $\sigma^z_{xy}$ of 1T'-MoS$_2$ is constant for Fermi energy within the fundamental gap. 
However, the absolute values of $\sigma^z_{xy}=-0.018$ (MoS$_2$) and $0.044~e^2/h$ (WS$_2$) for both 1T'-TMDCs are much
smaller than the conductance quantum $e^2/h$. 
The sign can be changed by rotating the Cartesian coordinate system appropriately.
Due to the rectangular symmetry of these 2D systems, the directions $x$ and $y$ are not equivalent and the exchange of the electric field and spin current directions yields different values $\sigma^z_{yx}=0.13$ (MoS$_2$) and $0.41~e^2/h$ (WS$_2$).
Neither the $C_{2h}$ point group contains mirror operations, in particular in-plane mirror symmetry in contrast to the hexagonal and square crystals, nor a classification of spin in up and down is possible.

For rectangular TIs our results drastically illustrate that a QSH phase of a 2D TI is not generally characterized by a quantized SH conductivity $\sigma^z_{xy}=\pm e^2/h$. 
This is only guaranteed, where the SOI operator commutes with the $z$-component $S_z$ of the electron spin. 
In this case time reversal flips both spin $S_z$ and $\sigma^z_{xy}$. 
If the electric field is applied in $y$-direction, the up and down spins give rise to Hall currents in opposite
$x$-directions. 
There is a net quantized SH conductivity $\sigma^{sH}_{xy}=\sigma^z_{xy}-\sigma^{-z}_{xy}=2e^2/h$
\cite{Hasan2010,Kane2005,Bernevig2006}. 
However, the spin conservation as well as the quantization of $\sigma^z_{xy}$ are violated by the Rashba-like SOI, which explicitly destroys the $z\rightarrow-z$ mirror symmetry, or contributions to the Hamiltonian, which
couple $\pi$ and $\sigma$ orbitals in the 2D system \cite{Kane2005a}. 
In complete electronic structure calculations,  spin conservation $[H_{SOI},S_z]=0$ does generally not occur. Considering that the electron momentum \textbf{\textit{p}} is confined in the $xy$-plane, with the Rashba SOI Hamiltonian $H_{SOI}$ the commutator yields $\sim E_z(p_yS_x-p_xS_y)$ for the commutator with the local electric field component $E_z$. Its expectation values vanish not only for the trivial situation of zero in-plane spin but also for high in-plane symmetry. The 2D projections of the studied hexagonal crystals effectively show sixfold rotational symmetry, which results in in-plane ``isotropy'' with $S_z$ conservation. In the rectangular cases with twofold rotations but no horizontal mirror operation in $C_{2h}$ the directions $x$ and $y$ are inequivalent, so that $S_z$ is not conserved and $\sigma_{xy}^z$ is far from the quantized value. The understanding of the near quantization in the $D_{4h}$ cases is somewhat more subtle, since $x$ and $y$ are symmetry-equivalent. Nevertheless, in real square 2D TIs $\sigma_{xy}^z$ deviates from the quantum $e^2/h$ by about 10\%. Despite non-quantized  spin Hall conductance $\sigma^z_{xy}$ a non-zero spin accumulation persists for the QSH phase, justifying the term quantum (but not quantized) SH effect, as indicated by $Z_2=1$\cite{Kane2005a}. 

\subsection{Influence of External Perturbations}

\begin{figure}[h]
	\includegraphics[width=\linewidth]{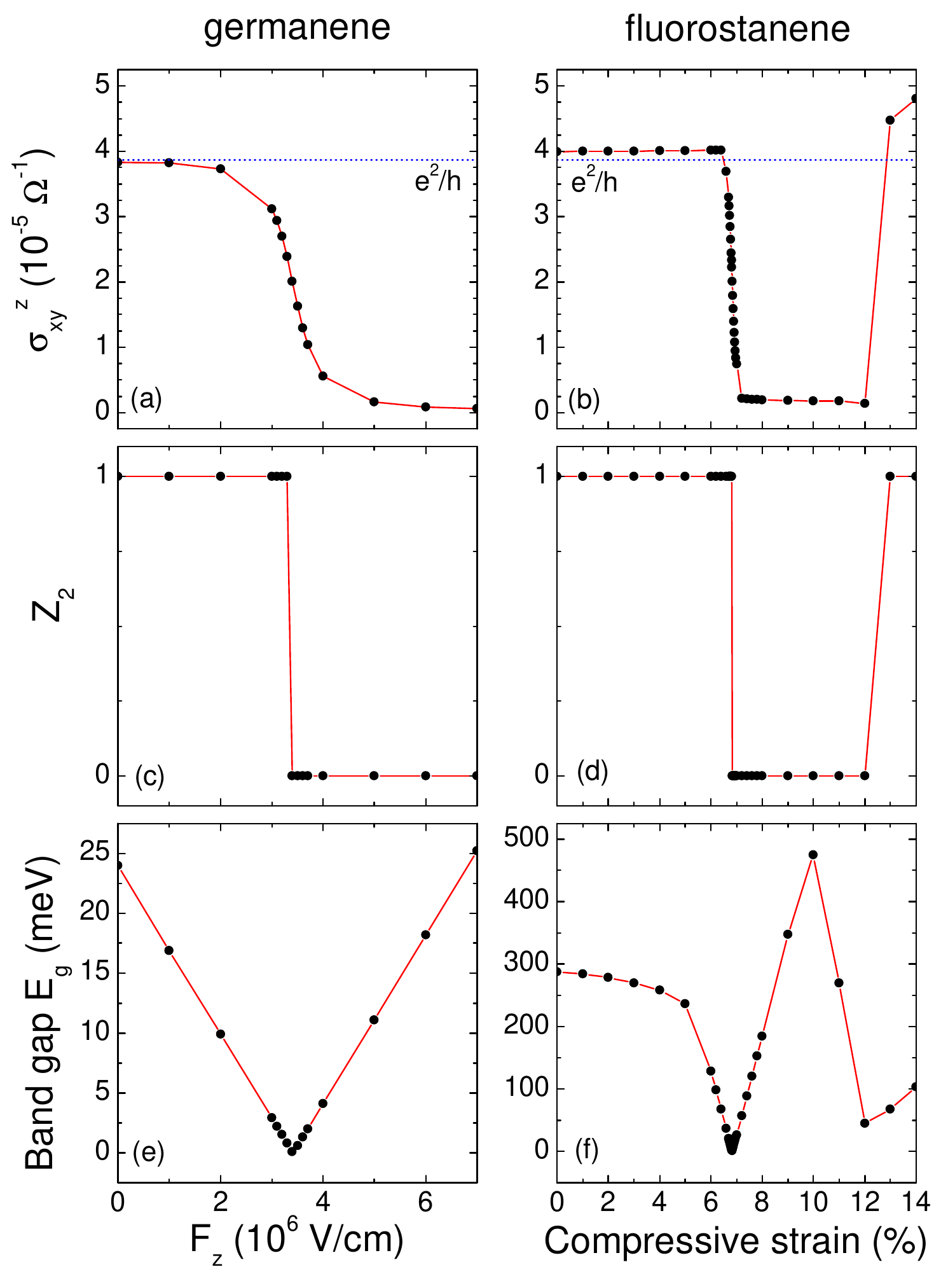} 
	\caption{Static spin Hall conductivity, $Z_2$ invariant and band gap in function of the applied electric field for germanene and as function of the compressive biaxial strain for fluorostanene. The blue dotted horizontal line illustrates the quantization value $e^2/h$.}
	\label{fig3}
\end{figure}

Electronic phase transitions and changes of the band topology induced by external perturbations
\cite{Matthes2014,Gao2017} will also influence the static SH conductivity
$\sigma^z_{xy}$, even for hexagonal 2D crystals.
In Fig.~\ref{fig3} we display the variation of the static spin Hall conductivity $\sigma^z_{xy}$, the $Z_2$ invariant, and the band gap $E_g$ as function of an external vertical electric field $F_z$ applied to germanene and of compressive biaxial strain on fluorostanene. Such a field breaks the mirror symmetry and modifies the SOI \cite{Acosta2019}.

The electrically tunable gap in the 2D TI germanene in Fig.~\ref{fig3}e is well known for this material but also for other group-IV honeycomb crystals
\cite{Liu2011a,Matthes2014,Ezawa2012,Drummond2012,Tsai2013} and even 2D topological crystalline insulators \cite{Liu2014a}. 
Extending the graph to negative field strengths $F_z$, the band gap $E_g$ shows a \textit{W} shape as a function of the external field with a gap closing at a critical field strength $F_{\rm crit}=3.3$$\times$10$^6$~V/cm (Fig.~\ref{fig3}c).
This behavior can be explained by the electric field modification of the Rashba-like SOI \cite{Matthes2014}.
The SOI modification leads to the lift of the band gap inversion and, consequently, tends toward a trivial band insulator.
The topological invariant $Z_2$ takes the values $Z_2=1$ or 0, indicating a topological phase transition along the electric field strength. 
The manifestation of the topological phase transition with the field in the static SH conductivity $\sigma^z_{xy}$ is described in Fig.~\ref{fig3}a. Away from the critical field strength $F_{\rm crit}$ the expected behavior for vanishing and large field strengths $F_z$ is visible. 
For zero field, i.e., for inversion-symmetric unbiased germanene, one observes the $e^2/h$ quantization. 
In the opposite limit of strong external fields, i.e., a trivial system without inversion symmetry and a non-inverted gap $E_g>20$~meV at the corner points $K$ and $K$' of the 2D hexagonal BZ, the static conductivity $\sigma^z_{xy}$ approaches the zero value.

Most surprising is the monotonous decrease from $\sigma^z_{xy}=e^2/h$ to $\sigma^z_{xy}=0$ with rising field strength, which is in contrast to the abrupt behavior of $Z_2$. 
The continuous transition is not directly related to the gap behavior and, hence, the Kane-Mele SOI
\cite{Matthes2014,Ezawa2012,Drummond2012,Tsai2013}. 
While the quantization does not depend on the SOI strength, rather on the symmetry, $\sigma^z_{xy}$ is modified with the strength of the perturbation in general.
When $F_z$ is applied to a buckled geometry with two sublattices, in which the two atoms within a hexagonal
unit cell are not coplanar, the inversion-symmetry break causes a non-vanishing staggered sublattice potential. 
The accompanying inversion-symmetry break from $D_{3d}$ to $C_{3v}$ explains the strict violation of the quantization with variation of $\sigma^z_{xy}$ versus the strength of the perturbing electric field. Consequently, the inversion symmetry is a key element for near quantization of $\sigma^z_{xy}$ even in hexagonal 2D TIs.

\begin{figure}[h]
	\includegraphics[width=\linewidth]{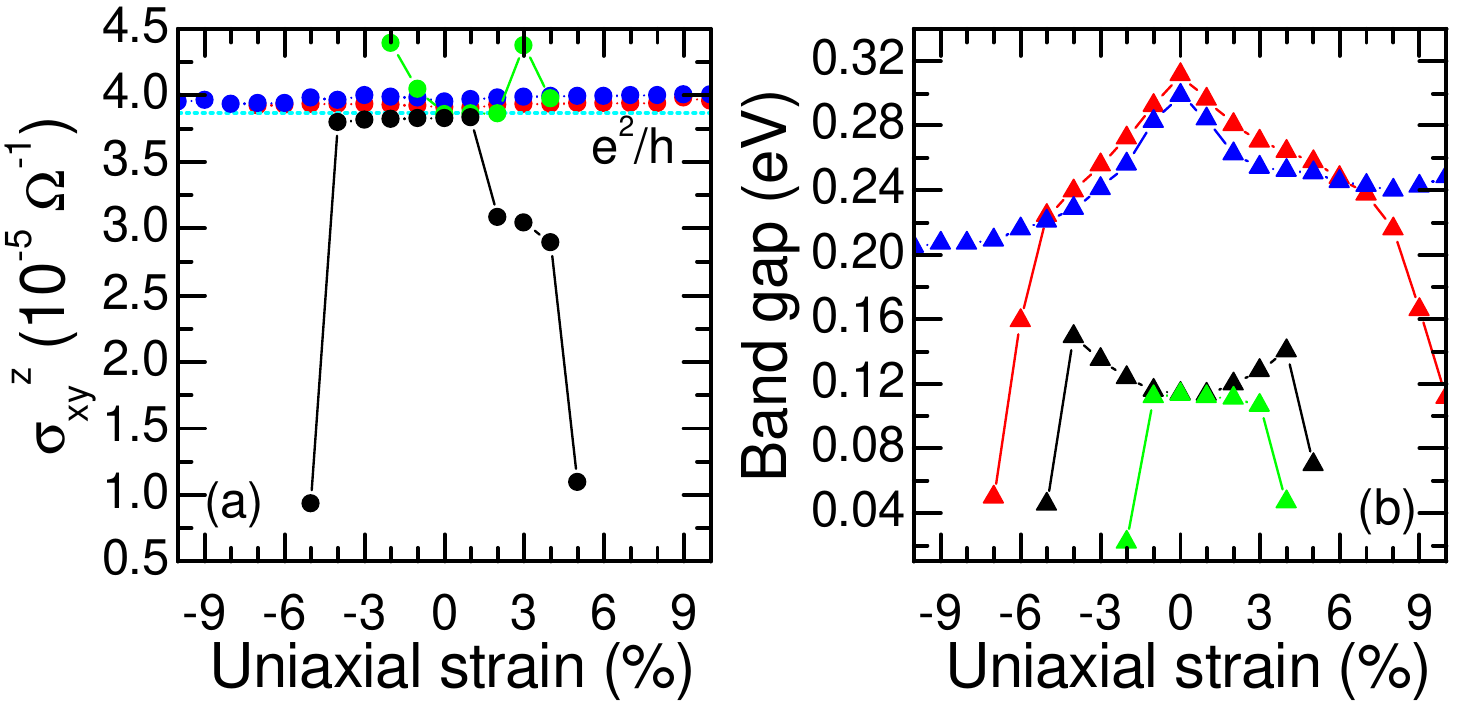} 
	\caption{Variation of static spin Hall conductivity (a) and band gap (b) as a function of applied uniaxial strain for germanene, stanene, GeI and SnF hexagonal crystals. The cyan dotted horizontal line illustrates the quantization value $e^2/h$. }
	\label{fig4}
\end{figure}

Biaxial strain is present if growing 2D crystals on a given substrate \cite{Matusalem2016,Matusalem2017}. 
It influences the electronic and optical properties of clean and functionalized group-IV materials such as stanene and
fluorostanene \cite{Modarresi2015,Lu2017a}. The influence of compressive biaxial
strain on fluorostanene is investigated in Figs.~\ref{fig3}b, d and f. 
With rising biaxial strain the fundamental inverted gap at $\Gamma$ of about 0.3~eV is closed to zero down to a reduction of the lattice constant of 7~\%.
The gap variation is accompanied by a topological phase transition as indicated by the abrupt change of both quantities, $Z_2$ and $\sigma^z_{xy}$, in Figs.~\ref{fig3}b and d, despite keeping the $D_{3d}$ symmetry. 
Further increase of the compressive strain leads to further band crossings.
For strain of about 12~\% again a gap closing together with an exchange of bands with opposite parities happens.
Consequently a high-strain-induced topological phase appears as displayed by $Z_2$ but qualitatively also indicated by $\sigma^z_{xy}$ with a near recovering of the quantized value. We also investigated non-symmetry conserving compressive and tensile uniaxial strains in the basal plane of the pure and halogenated group-IV materials germanene, stanene, GeI and SnF in Fig. \ref{fig4}. Applying uniaxial strain of $\pm$ 2\% we do not found significant variations of $\sigma_{xy}^z$ and the band gap. Such a weak symmetry lowering has only a minor influence. Only for strains of about $\pm$ 4\% or bigger significant gap reductions and modifications of the spin Hall conductivity occur. 

\subsection{Model Topological Insulator}

For a better understanding of the influence of symmetry, external perturbations and spin-orbit coupling we also investigate the static spin Hall conductivity for model systems. As example, we study low-buckled honeycomb crystals under the influence of vertical electric fields and/or in-plane uniaxial strain. They could be derived from group-IV materials silicene, germanene and stanene. Their electronic properties are ruled by nearly linear bands around the corner points $K$ and $K'$ of the hexagonal BZ, the Dirac touching points. SOI opens a small gap so that for extremely small deviations $\boldsymbol{\kappa}$ in $\boldsymbol{k}$-space from $K$ or $K'$ massive Dirac fermions appear\cite{Matthes2013}. In a tight-binding (TB) approximation the four relevant bands near $K$ can be simulated by a $4\times 4$ Hamiltonian\cite{Liu2011,Dyrda2012}

{\footnotesize
	\begin{equation}\label{eq3}
	\hat{H}(\boldsymbol{\kappa})=
	\begin{pmatrix}
	\hat{h}(\boldsymbol{\kappa}) & \hbar(v_{F_x}\kappa_x + iv_{F_y}\kappa_y)\hat{\sigma}_0 \\
	\hbar(v_{F_x}\kappa_x - iv_{F_y}\kappa_y)\hat{\sigma}_0 & -\hat{h}(\boldsymbol{\kappa})
	\end{pmatrix}
	\end{equation}}
with the $2\times 2$ diagonal blocks

{\footnotesize
	\begin{equation}\label{eq4}
	\hat{h}(\boldsymbol{\kappa})=-\lambda_{SO}\hat{\sigma_z}-a\lambda_R(\kappa_y\hat{\sigma_x}-\kappa_x\hat{\sigma_y})+U\hat{\sigma_0}
	\end{equation}}\noindent where the unit matrix $\hat{\sigma}_0$ and the Pauli matrices $\hat{\sigma_{\alpha}}, (\alpha=x,y,z)$ appear. The difference of the Fermi velocities  $v_{F_x}$ and $v_{F_y}$ characterizes a small anisotropy induced by an external perturbation. SOI is characterized by two constants $\lambda_{SO}$ and $\lambda_R$. While $2\lambda_{SO}$ determines the fundamental gap, $\lambda_R$ rules the Rashba splitting. The gate voltage $U$ models a vertical electric field. 

Details of the electronic-structure model and the calculation of the static spin Hall conductivity can be found in the \textbf{Appendix}. In the low-temperature limit it results ($\lambda_R>0$) 

{\footnotesize
	\begin{widetext}
		\begin{eqnarray}\label{eq5}
		\sigma^z_{xy}=\frac{e^2}{h}\frac{v_{F_x}v_{F_y}}{\sqrt{v_{F_x}^2+a^2\lambda_R^2/\hbar^2}\sqrt{v_{F_y}^2+a^2\lambda_R^2/\hbar^2}}\left\{\Theta(||U|-\lambda_{SO}|-|\mu|)+\frac{1}{2}\left[1-\frac{|U|-\lambda_{SO}}{|\mu|}\right]\Theta(|U|+\lambda_{SO}-|\mu|)\Theta(|\mu|-||U|-\lambda_{SO}|)+\right.\nonumber\\
		\left. \frac{\lambda_{SO}}{|\mu|}\Theta(|\mu|-|U|-\lambda_{SO})\right\} 
		\end{eqnarray}
\end{widetext}}\noindent from (\ref{eqm8}). Thereby, the chemical potential $\mu$ is measured with respect to the midgap position and can vary from positions in the conduction bands up to those in the valence bands. 

Expression (\ref{eq5}) illustrates the central role of SOI for the quantization of the spin Hall conductance and its explicit size. This especially holds for the Rashba contribution to the SOI. In the limit $\lambda_R\rightarrow 0$, even the anisotropy of the Dirac cones does not play any role. 

Without external pertubations in (\ref{eq5}), i.e., $v_{F_x}=v_{F_y}=v_F$ and $U=0$, one has

{\footnotesize
	\begin{eqnarray}\label{eqm9}
	\sigma^z_{xy}=\frac{e^2}{h}\frac{1}{1+(a\lambda_R)^2/(\hbar v_F)^2}\bigg[\Theta(\lambda_{SO}-|\mu|)+ \nonumber \\
	\frac{\lambda_{SO}}{|\mu|}\Theta(|\mu|-\lambda_{SO})\bigg].
	\end{eqnarray}}\noindent One can conclude that for Fermi level in the gap, $\lambda_{SO}>|\mu|$, the spin Hall conductivity is generally not quantized. This is due to the Rashba SOI ruled by the parameter $\lambda_R$. It violates the conservation of the $z$ component $S_z$ of the spin. Using $\hat{S}_z$ in TB representation, it holds

{\footnotesize
	\begin{equation}\label{eqm10}
	[\hat{S}_z,\hat{H}(\boldsymbol{\kappa})]=i\hbar a\lambda_R
	\begin{pmatrix}
	0 & -\kappa_x+i\kappa_y & 0 & \\
	-\kappa_x-i\kappa_y & 0 & 0 & 0 \\
	0 & 0 & 0 & \kappa_x-i\kappa_y \\
	0 & 0 & \kappa_x+i\kappa_y & 0 \\
	\end{pmatrix}.
	\end{equation}}\noindent Conservation violation is proportional to the deviation $\boldsymbol{\kappa}$. Only directly at a $K$ or $K'$ point the spin is conserved. However, the parameter $\lambda_R$ is relatively small in silicene, germanene and stanene\cite{Liu2011}. The term $(a\lambda_R)^2/(\hbar v_F)^2$ is of the order of $2\times 10^{-4}$ or even much smaller in silicene. Thus, the influence of the Rashba SOI on the spin Hall conductivity is negligible. This is the reason for the seemingly quantization observed in Figs. 1(a) and (c). This conclusion is in line with predictions in literature\cite{Liu2011,Ezawa2012a}. The negligible influence of $\lambda_R$ also explains that deformations of the Dirac cones, here described by $v_{F_x}\neq v_{F_y}$, do not play a role. Indeed, in Fig. \ref{fig4} in-plane uniaxial strain up to $\pm 2\%$ is found not to influence the quantization of $\sigma_{xy}^z$ in germanene with $\mu$ in a midgap position. In the isotropic limit and vanishing $\lambda_R$ the prefactor in (\ref{eq5}) also becomes the conductance quantum $e^2/h$. Expression (\ref{eq5}) does not explain the field dependence in Fig. 3(a) for germanene and $\mu=0$, because of the neglect of the interplay of SOI and external field.

\section{Summary and Conclusions}

Using \emph{ab initio} electronic-structure calculations for energy eigenvalues and eigenfunctions including spin-orbit interaction, we have investigated the third-rank tensor of the static spin Hall conductivity for 2D crystals with various geometries. 
Central point of our studies was the conductivity quantization $e^2/h$ and the effect of spin-orbit interaction. The influence of the latter one is also studied applying a model Hamiltonian. 
The findings confirm the conclusion that a quantum spin Hall phase is not generally characterized by a quantized spin Hall conductivity.
Rather, the quantization significantly depends on the atomic geometry of the topological insulators but also its SOI.

For hexagonal honeycomb crystals with inversion symmetry, systems with high symmetry and very small Rashba SOI contribution, the quantization has been almost numerically demonstrated. 
The quantization $e^2/h$ is completely violated for rectangular crystals and normal spin components. Even, for square crystals the results deviate significantly from $e^2/h$.
In addition, we have shown that external perturbations, such as vertical electric fields and compressive biaxial strains, induce topological phase transitions in honeycomb sheet crystals with abrupt changes of the topological invariant between $Z_2=1$ or 0 but smooth or abrupt changes of the spin Hall conductivity between $e^2/h$ and 0
in dependence on inversion symmetry or not. 

As a rule of thumb we found that a quantized spin Hall conductance only appears for inversion-symmetric 2D TIs with almost in-plane isotropy. There, the effect of the Rashba SOI remains weak, even for not too small coupling constants. Lifting either the hexagonal or the inversion symmetry destroys the near quantization. In-plane mirror symmetry is a necessary (see hexagonal crystals) but not sufficient (see rectangular TIs) condition for near quantization. Even in the high-symmetry TIs a weak or better almost vanishing Rashba contribution to SOI is needed to measure a static spin Hall conductance close to the conductance quantum.

\begin{acknowledgments}
This work was supported by the Brazilian funding agencies CNPq (Grants. No. 477366/2013-9 Universal), CAPES (grants No. 23038.005810/2014-34 and No. 88881.068355/2014-01 within the PVE/CsF program and scholarship grant), and S\~ao Paulo Research Foundation (FAPESP), (grants No. 2012/507383-3 and No. 2014/13907-7).
\end{acknowledgments}

\appendix*
\section{Model Studies}

We investigate the low-buckled, graphene-like group-IV 2D crystals silicene, germanene, and stanene as model systems. In their case the static spin Hall conductivity $\sigma^z_{xy}$ should be mainly influenced but the two lowest unoccupied and the two highest occupied bands near the $K$ and $K'$ points of the BZ, because of the vanishing interband energies $\hbar\omega\rightarrow 0$, which contribute to the conductivity values. This fact is illustrated in Figs. S3(a), S4(a), and S7, which also show a two-fold band degeneracy without external pertubation. The interesting parts of the band structure around $K$ and $K'$ points can be modeled in the framework of a tight-binding (TB) approach\cite{Liu2011}. If instead of a Bloch basis a basis of non-overlapping localized orbitals $\rvert\alpha\rangle$ is used, the static spin Hall conductivity (\ref{eqm1}) can be rewritten in a general form\cite{Matthes2016a}

{\footnotesize
	\begin{widetext}
		\begin{equation}
		\sigma^z_{xy} = \lim_{\omega\rightarrow 0}\frac{e^2}{i\omega}\frac{1}{A}\sum_{\mathbf{k}}\int_{-\infty}^{\infty}\frac{d\varepsilon}{2\pi}\int_{-\infty}^{\infty}\frac{d\varepsilon'}{2\pi}\frac{f(\varepsilon)-f(\varepsilon')}{\varepsilon-\varepsilon' + \hbar\omega + i\eta}Tr\bigg\{[\hat{V}^z_x\hat{A}(\mathbf{k},\varepsilon')][\hat{V}_y\hat{A}(\mathbf{k},\varepsilon)]\bigg\} \label{eqm1}
		\end{equation}
\end{widetext}}
\noindent with the matrices of the spin current velocity $\hat{V}_x^z$ and charge current velocity $\hat{V}_y$ in the localized basis. The matrix of the spectral function $\hat{A}(\mathbf{k},\varepsilon)$ at a certain point in the BZ is derived from the single-particle Green function $\hat{G}(\mathbf{k},\hbar z)$ as 

{\footnotesize
	\begin{equation}\label{eqm2}
	\hat{G}(\mathbf{k},\hbar z) = \int_{-\infty}^{\infty}\frac{d\varepsilon}{2\pi}\frac{\hat{A}(\mathbf{k},\varepsilon)}{\hbar z - \varepsilon}. 
	\end{equation}}\noindent Using the complex energies $\hbar z = \hbar\omega \pm i\eta$ $(\eta\rightarrow +0)$ the spectral function can be directly expressed by a difference of Green functions. 

Important contributions to the spin Hall conductivity only arise from $\mathbf{k}$-point regions around $K$ and $K'$ points. Therefore, we study in the average one $K$ and one $K'$ point region and replace the wavevector $\mathbf{k}$ by the portion of $K$ or $K'$ and some deviations $\boldsymbol{\kappa}$. In the sublattice space, the matrix form of the corresponding $4\times 4$ Hamiltonian $\hat{H}(\boldsymbol{\kappa})$ for a $K$ point takes the form\cite{Liu2011,Dyrda2012}

{\footnotesize
	\begin{equation}\label{eqm3}
	\hat{H}(\boldsymbol{\kappa})=
	\begin{pmatrix}
	\hat{h}(\boldsymbol{\kappa}) & \hbar(v_{F_x}\kappa_x + iv_{F_y}\kappa_y)\hat{\sigma}_0 \\
	\hbar(v_{F_x}\kappa_x - iv_{F_y}\kappa_y)\hat{\sigma}_0 & -\hat{h}(\boldsymbol{\kappa})
	\end{pmatrix}
	\end{equation}}
with the $2\times 2$ diagonal parts

{\footnotesize
	\begin{equation}\label{eqm4}
	\hat{h}(\boldsymbol{\kappa})=-\lambda_{SO}\hat{\sigma_z}-a\lambda_R(\kappa_y\hat{\sigma_x}-\kappa_x\hat{\sigma_y})+U\hat{\sigma_0},
	\end{equation}}\noindent where $\hat{\sigma}_{\alpha}$ are the unit ($\alpha=0$) and Pauli ($\alpha=x,y,z$) matrices in the spin space are introduced. The $4\times 4$ character of the Hamiltonian (\ref{eqm3}) is due to the basis $\{|\phi_A\rangle,|\phi_B\rangle\}\otimes\{|\uparrow\rangle,|\downarrow\rangle\}$ of a $p$-orbital at an atom of sublattice A or B and the spin orbitals. First- and second-nearest-neighbor interaction have been taken into account. The Fermi velocities $v_{F_x}$ and $v_{F_y}$ characterizing the Dirac cones near the Dirac points $K$ and $K'$ are dominated by the nearest-neighbor hopping integrals. Despite the buckling of the group-IV basal plane the nearest-neighbor spin-orbit interaction vanishes, in contrast to the next-nearest neighbor one. The latter is divided into two parts, according to the two components of the internal electric field parallel with or perpendicular to the plane. The different spin components are coupled by the constants $\lambda_{SO}$ or $\lambda_R$ in (\ref{eqm4}). The contribution $-a\lambda_R(\kappa_y\hat{\sigma_x}-\kappa_x\hat{\sigma_y})$ with the lattice parameter $a$ clearly characterizes the Rashba contribution to SOI\cite{Bychkov1984}, while the second intrinsic contribution $-\lambda_{SO}\hat{\sigma_z}$ is responsible for the opening of a small gap at $K$ and $K'$ and the finite mass of Dirac particles\cite{Matthes2013}. 

We have to mention that around a Dirac point $K'$ the Hamiltonian (\ref{eqm3}) becomes within the same localized basis

{\footnotesize
	\begin{equation}\label{eqm5}
	\hat{H}(\boldsymbol{\kappa})=
	\begin{pmatrix}
	-\hat{h}(\boldsymbol{\kappa}) & \hbar(v_{F_x}\kappa_x - iv_{F_y}\kappa_y)\hat{\sigma}_0 \\
	\hbar(v_{F_x}\kappa_x + iv_{F_y}\kappa_y)\hat{\sigma}_0 & \hat{h}(\boldsymbol{\kappa})
	\end{pmatrix}.
	\end{equation}}\noindent However, it yields the same contribution to the spin Hall conductivity as $K$. In comparison to the common TB Hamiltonian for silicene and similar materials\cite{Liu2011}, two generalizations are taken into account in (\ref{eqm5}). We allow for the application of an external vertical electric field characterized by a gate voltage $U$ (in energy units). A weak in-plane anisotropy is modeled by a deformation of the Dirac cones as illustrated by different Fermi velocities $v_{F_x}$ and $v_{F_y}$. Their difference may be induced by uniaxial in-plane or an anisotropic biaxial strain. 

A Hamiltonian of the type (\ref{eqm3}) or (\ref{eqm5}) can be easily diagonalized. As a result one finds the four bands around a Dirac point ($\kappa^2=\kappa_x^2+\kappa_y^2$)

{\footnotesize
	\begin{eqnarray}\label{eqm6}
	\varepsilon_{\pm}(\boldsymbol{\kappa})=\pm\bigg[\lambda_{SO}^2+U^2+\hbar^2(v_{F_x}^2\kappa_x^2+v_{F_y}^2\kappa_y^2) +\\
	a^2\lambda_R^2\kappa^2\pm2|U|\sqrt{\lambda_{SO}^2+a^2\lambda_R^2\kappa^2}\bigg]^{1/2}. \nonumber
	\end{eqnarray}}\noindent While the two conduction (valence) bands are degenerate for $U=0$ (see Figs. S3(a) and S4(a)), they split for a finite gate voltage as indicated in the insets of Figs. S7(b) and (c). There is a fundamental gap in the band structure at $K$ or $K'$ of $2(\lambda_{SO}-|U|)$. For small electric fields $\lambda_{SO} > |U|$ the systems are topological insulators\cite{Liu2011a}. Above the critical field strengths,  $|U|t > \lambda_{SO}$ silicene, germanene and stanene become trivial insulators\cite{Ezawa2012a,Matthes2014}. 

The TB Hamiltonian (\ref{eqm3}) allows for the direct calculation of the velocity matrices in expression (\ref{eqm1}) as 

{\footnotesize
	\begin{equation*}
	\hat{V}_x=\frac{1}{\hbar}\frac{\partial}{\partial \kappa_x}\hat{H}(\boldsymbol{\kappa})=\frac{1}{\hbar}
	\begin{pmatrix}
	0 & -ia\lambda_R & \hbar v_{F_x} & 0 \\
	ia\lambda_R & 0 & 0 & \hbar v_{F_x} \\
	\hbar v_{F_x} & 0 & 0 & ia\lambda_R \\
	0 & \hbar v_{F_x} & -ia\lambda_R & 0 
	\end{pmatrix},
	\end{equation*}}

{\footnotesize
	\begin{equation}\label{eqm7}
	\hat{V}_y=\frac{1}{\hbar}\frac{\partial}{\partial \kappa_y}\hat{H}(\boldsymbol{\kappa})=\frac{1}{\hbar}
	\begin{pmatrix}
	0 & -a\lambda_R & i\hbar v_{F_y} & 0 \\
	-a\lambda_R & 0 & 0 & i\hbar v_{F_y} \\
	-i\hbar v_{F_y} & 0 & 0 & a\lambda_R \\
	0 & -i\hbar v_{F_y} & a\lambda_R & 0 
	\end{pmatrix},
	\end{equation}}

{\footnotesize
	\begin{equation*}
	\hat{V}_x^z=\frac{1}{2\hbar}\left[\hat{V}_x,\hat{S}_z\right]_+=\frac{1}{\hbar}
	\begin{pmatrix}
	0 & 0 & \hbar v_{F_x} & 0 \\
	0 & 0 & 0 & -\hbar v_{F_x} \\
	\hbar v_{F_x} & 0 & 0 & 0 \\
	0 & -\hbar v_{F_x} & 0 & 0 
	\end{pmatrix}.
	\end{equation*}}\noindent The Green function matrix (\ref{eqm2}) also follows from the Hamiltonian (\ref{eqm3}) by inversion of the $4\times 4$ matrix $[\hat{H}(\boldsymbol{\kappa})-\hbar z]$. It directly determines the spectral function $\hat{A}(\boldsymbol{\kappa},\varepsilon)$. 

All these matrices are used to calculate the trace in (\ref{eqm1}). The Dirac $\delta$-functions at the band energies (\ref{eqm6}) are applied to perform the energy integration in (\ref{eqm1}). The BZ summation is divided into the contributions of $K$ and $K'$ with explicit integrations. Neglecting the additional influence of $\lambda_R$ due to the presence of a gate voltage $U$ the static spin Hall conductivity is written in the compact form in the low-temperature limit

{\footnotesize
	\begin{widetext}
		\begin{equation}\label{eqm8}
		\sigma^z_{xy}=\frac{e^2}{h}\frac{v_{F_x}v_{F_y}}{\sqrt{v_{F_x}^2+a^2\lambda_R^2/\hbar^2}\sqrt{v_{F_y}^2+a^2\lambda_R^2/\hbar^2}}\frac{1}{2}\sum_{\nu=1,2}\left\{\Theta(\lambda_{SO}+(-1)^{\nu}|U|-|\mu|)+\frac{\lambda_{SO}+(-1)^{\nu}|U|}{|\mu|}\Theta(|\mu|-\lambda_{SO}-(-1)^{\nu}|U|)\right\}
		\end{equation}
\end{widetext}}\noindent with the chemical potential $\mu$ of the electron measured with respect to the Dirac point energy $\varepsilon=0$.

%
\end{document}


\title{Supplemental Material for: Quantization of spin Hall conductivity in two-dimensional topological insulators versus symmetry and spin orbit interaction}
	\author{Filipe Matusalem}\email{filipematus@gmail.com, gmsn@ita.br}\affiliation{\gmsn}
\author{Lars Matthes}\affiliation{\jena}
\author{J\"urgen Furthm\"uller}\affiliation{\jena}
\author{Marcelo Marques}\affiliation{\gmsn}
\author{Lara K. Teles}\affiliation{\gmsn}
\author{Friedhelm Bechstedt}\affiliation{\jena}


\maketitle

\thispagestyle{empty}

\section{Atomic Geometries}

\begin{figure}[ht]
	\includegraphics[width=0.8\linewidth]{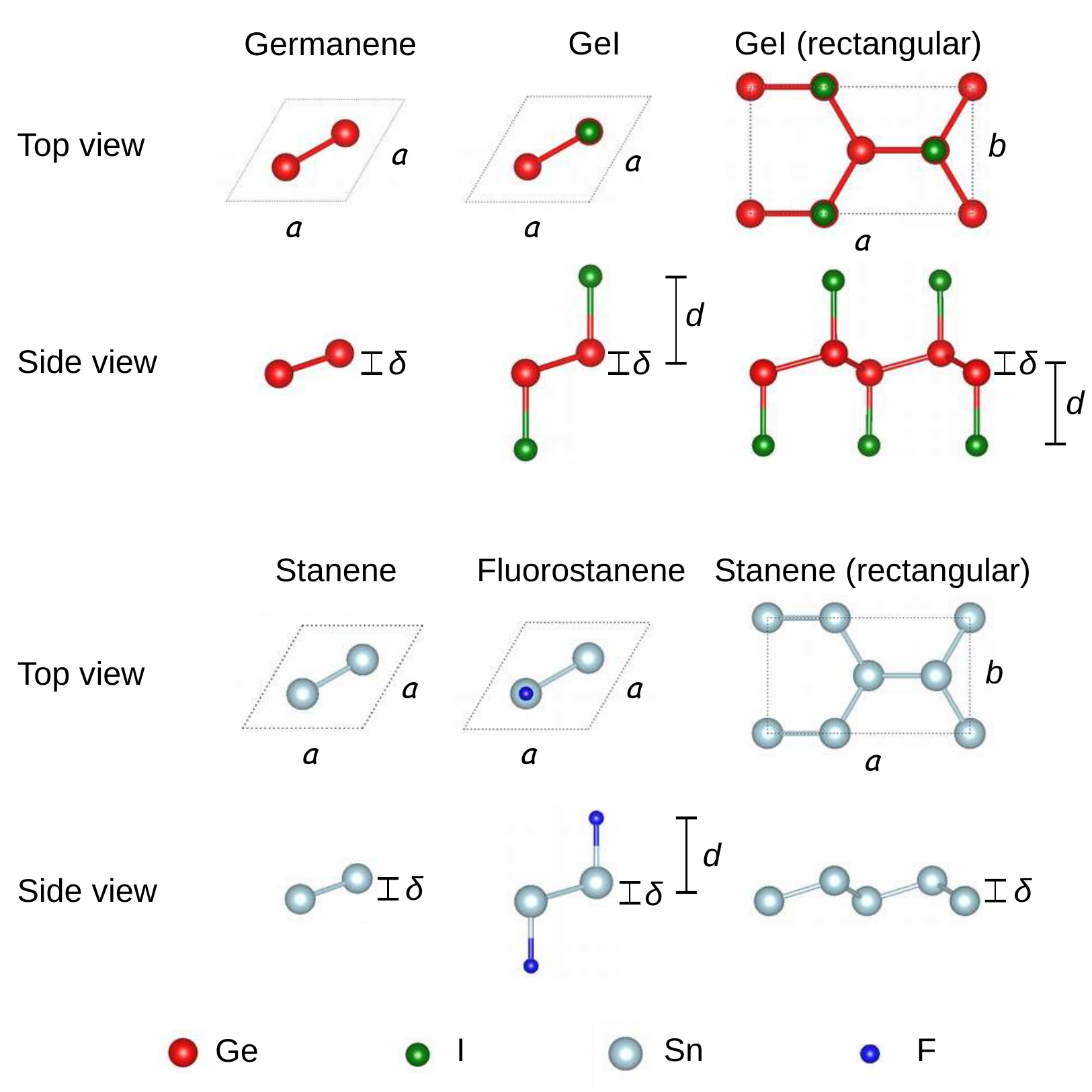}
	\caption{\label{hex} Atomic geometries in the unit cell (dashed line) of germanene, iodine-decorated germanene (GeI), stanene and fluorine-decorated stanene (fluorostanene). Besides a primitive hexagonal cell also a non-primitive rectangular one is displayed for GeI and stanene.}
\end{figure}

\begin{figure}[ht]
	\includegraphics[width=0.8\linewidth]{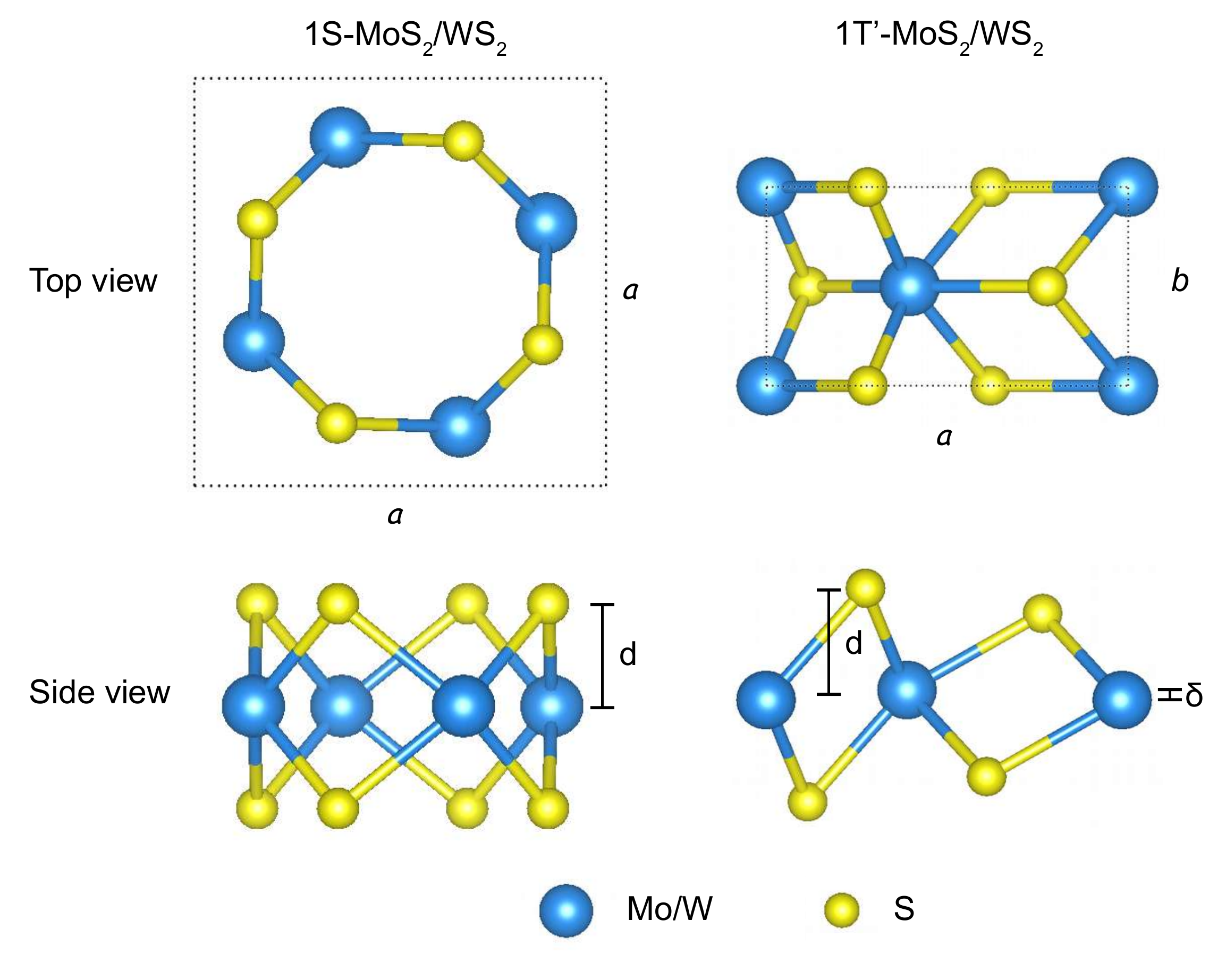}
	\caption{\label{mos2} Atomic geometries in square and rectangular unit cells (dashed line) of 1S-MoS$_2$/WS$_2$ and 1T'-MoS$_2$/WS$_2$.}
\end{figure}

\begin{table*}[ht]
	\centering
	\caption{\label{tab:s1} Atomic geometric parameters, as indicated in Figs. \ref{hex} and \ref{mos2}, for all systems under study. }
	\begin{tabular}{|c|c|c|c|c|c|c|c|c|c|c|}
		\hline
		         & germanene & GeI  & GeI (rectangular) & stanene & fluorostanene & stanene (rectangular) & 1S-MoS$_2$ & 1S-WS$_2$ & 1T'-MoS$_2$ & 1T'-WS$_2$ \\
		\hline
		  $a$    &   4.06    & 4.31 &       7.47        &  4.68   &     4.94      &         8.10          &    6.34    &   6.36    &    5.72     &    5.71    \\
		  $b$    &     -     &  -   &       4.31        &    -    &       -       &         4.68          &     -      &     -     &    3.17     &    3.19    \\
		$\delta$ &   0.69    & 0.69 &       0.69        &  0.85   &     0.72      &         0.85          &     0      &     0     &    0.15     &    0.14    \\
		   d     &     -     & 2.91 &       2.91        &    -    &     2.82      &           -           &    1.56    &   1.55    &    1.73     &    1.74    \\
		\hline
	\end{tabular}
\end{table*}

\section{Band Structures}

\begin{figure}[H]
	\centering
	\subfigure[{} germanene]{\includegraphics[scale=0.31]{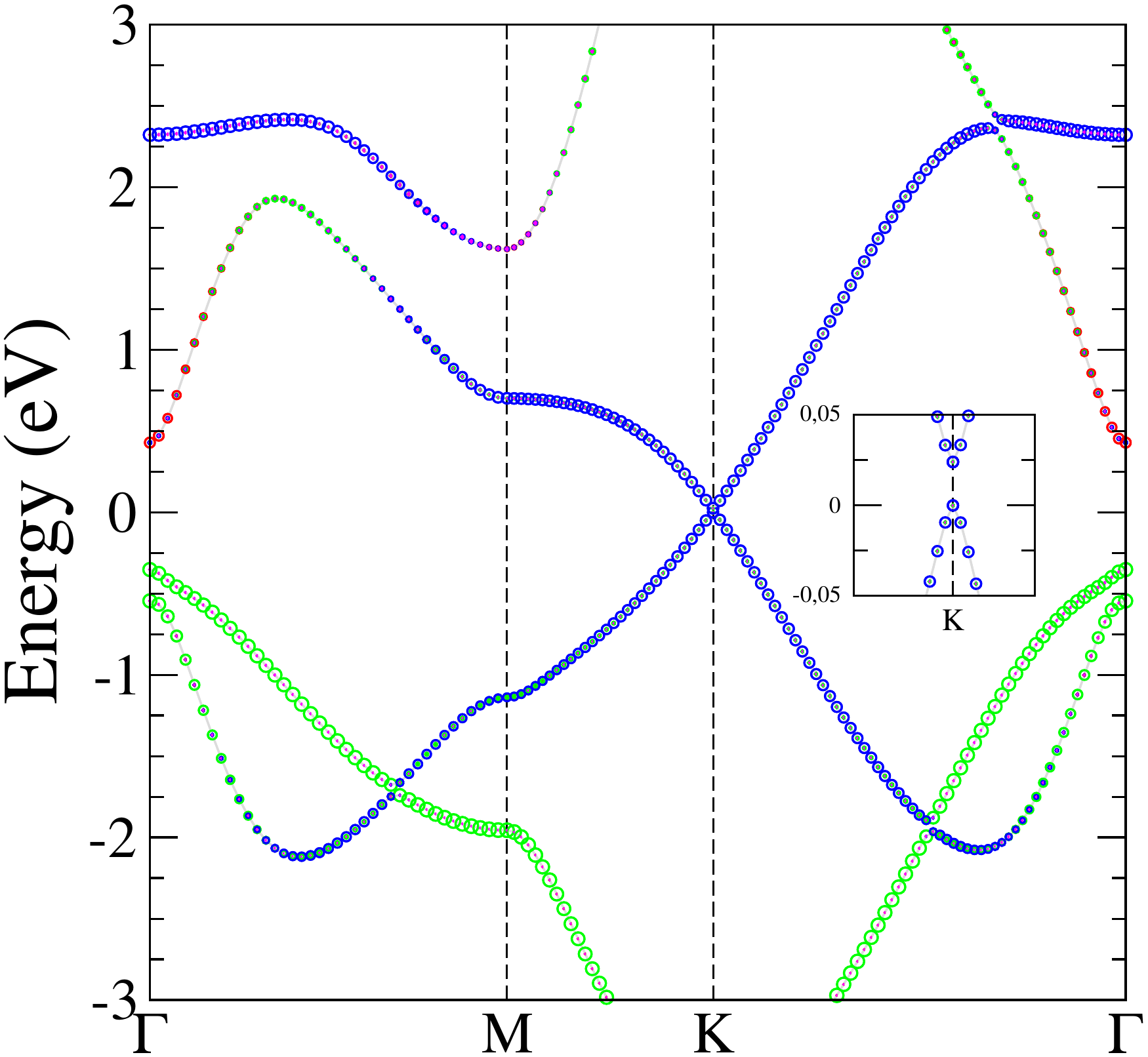}}\qquad
	\subfigure[{} GeI]{\includegraphics[scale=0.31]{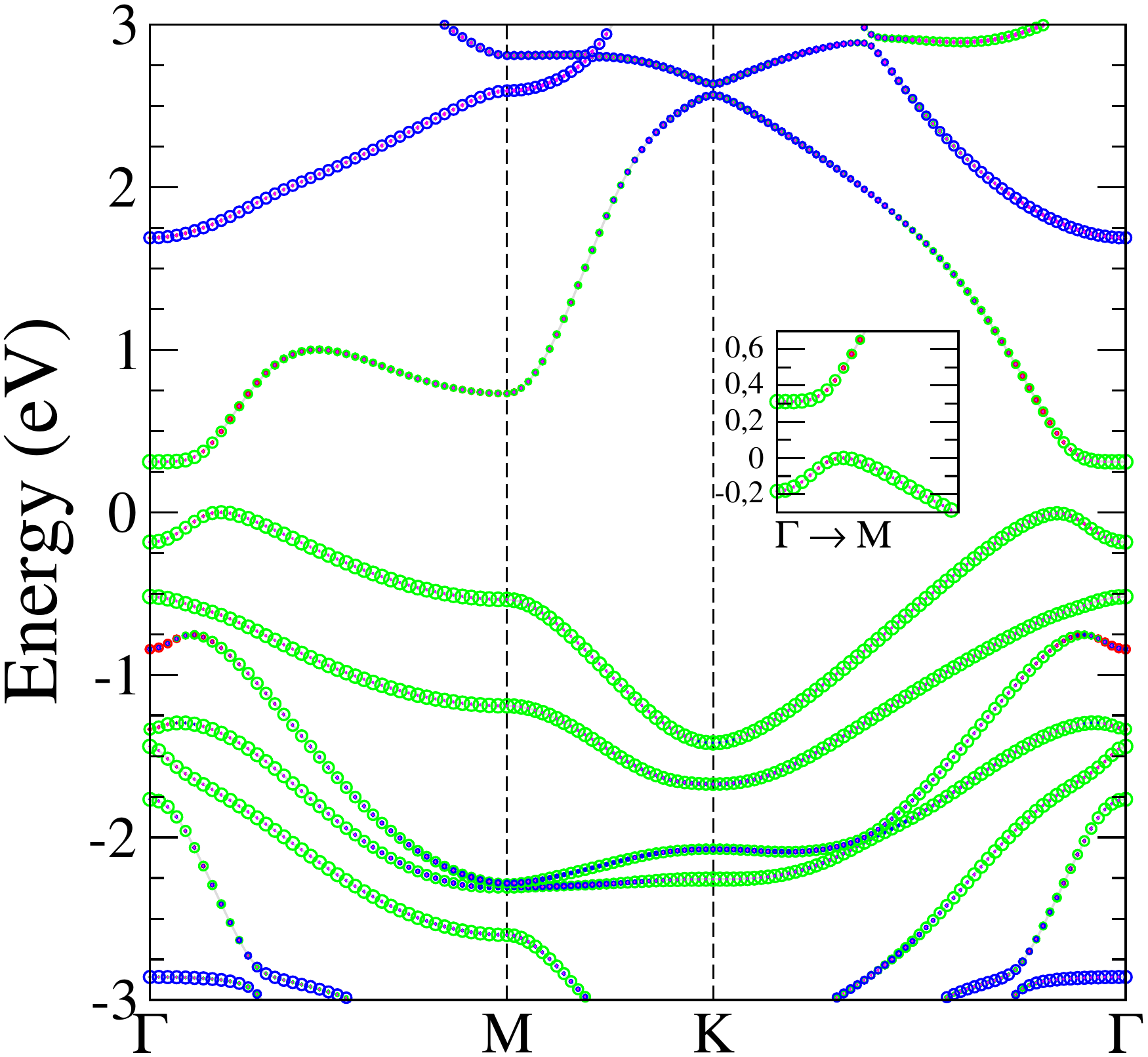}}\qquad
	\subfigure[{} GeI (rectangular)]{\includegraphics[scale=0.31]{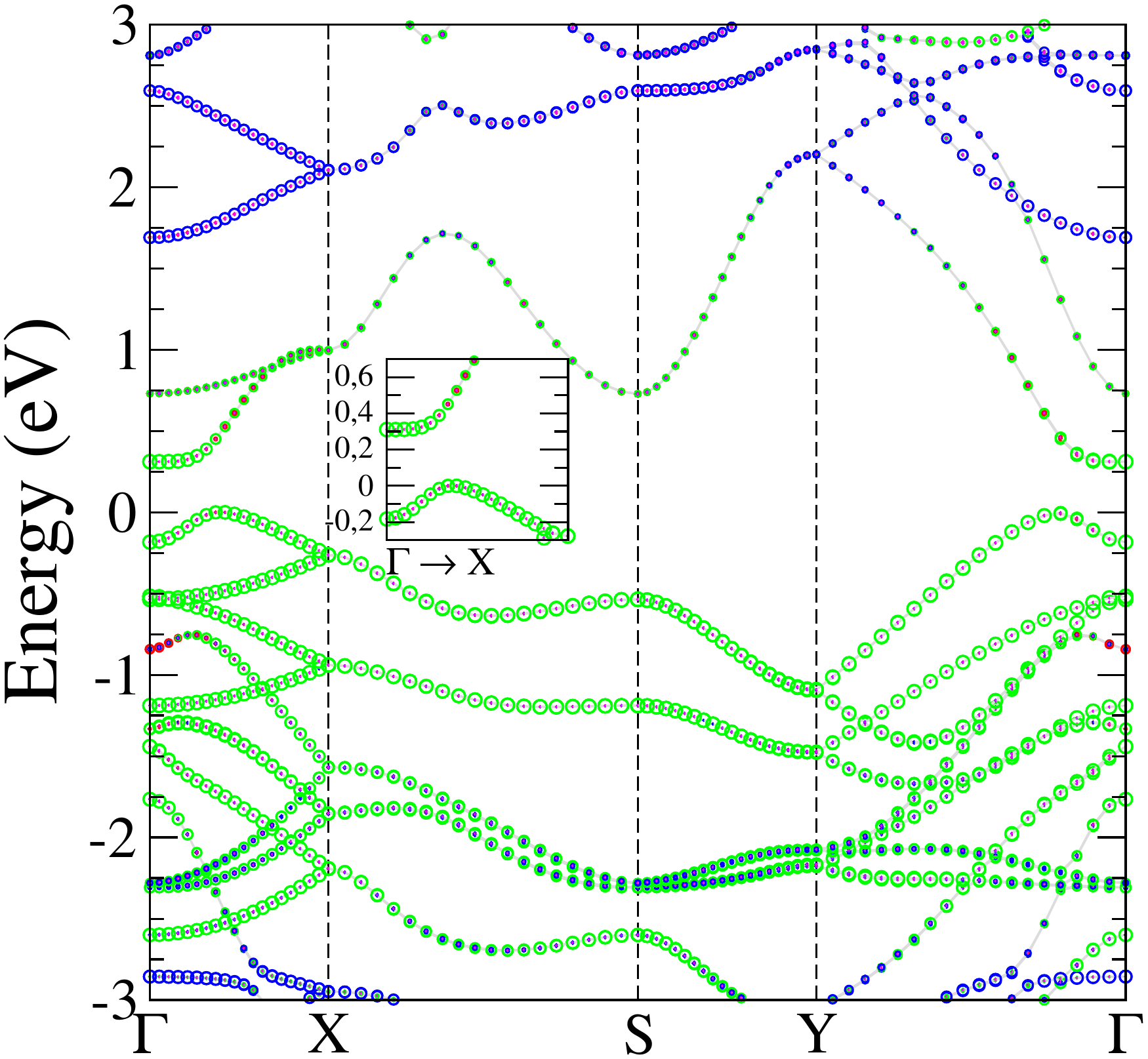}}

	\caption{\label{band-hex} Band structures for (a) germanene, (b) iodine-decorated germanene (hexagonal cell) and (c) iodine-decorated germanene (rectangular cell). Red, green, blue and magenta circles depict relative contributions from $s$, $p_x+p_y$ , $p_z$ and $d$ orbitals, respectively, to the band character. The insets display the gap regions. }
\end{figure}

\begin{figure}[H]
	\centering
	\subfigure[{} stanene]{\includegraphics[scale=0.31]{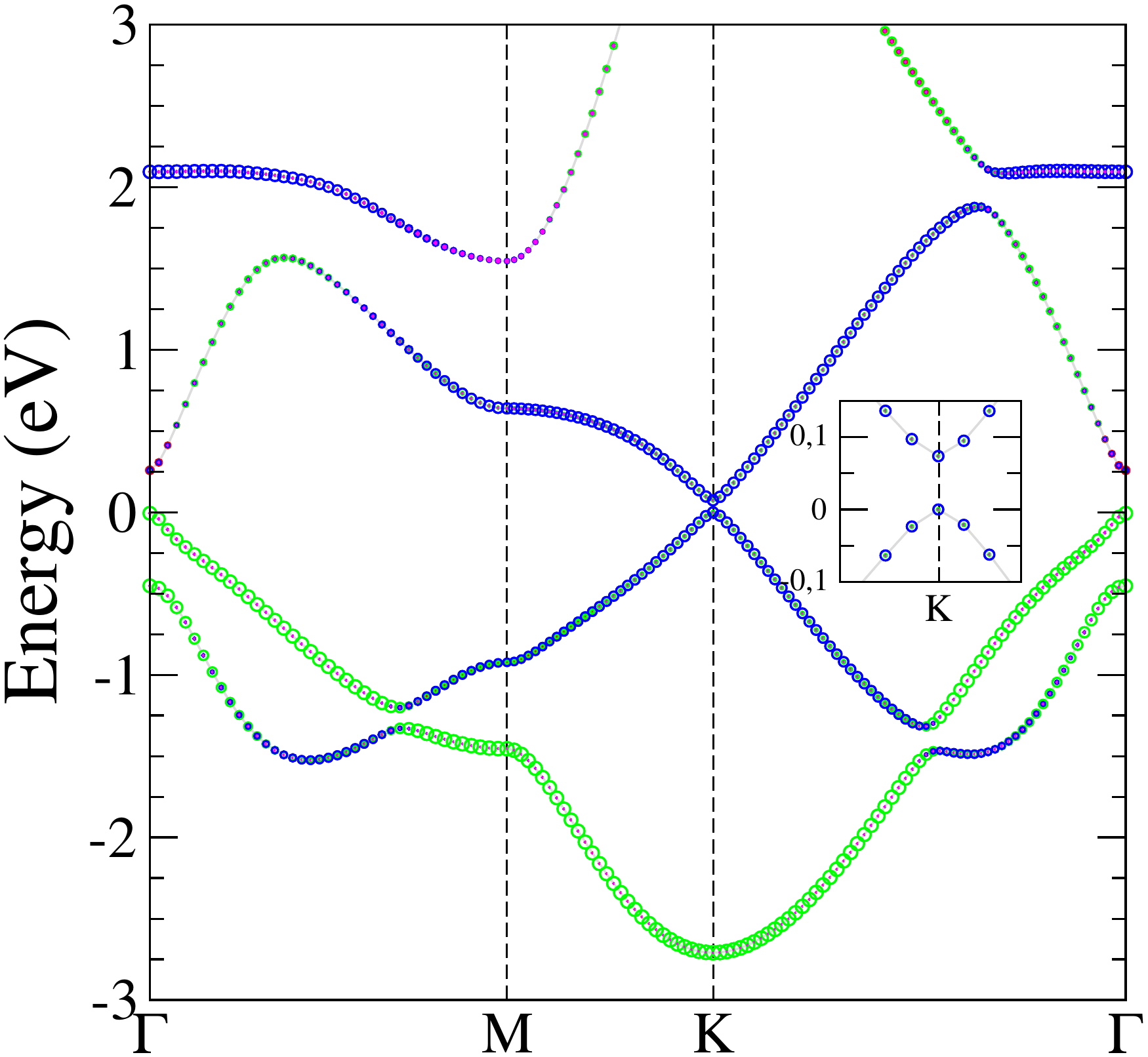}}\qquad
	\subfigure[{} fluorostanene]{\includegraphics[scale=0.31]{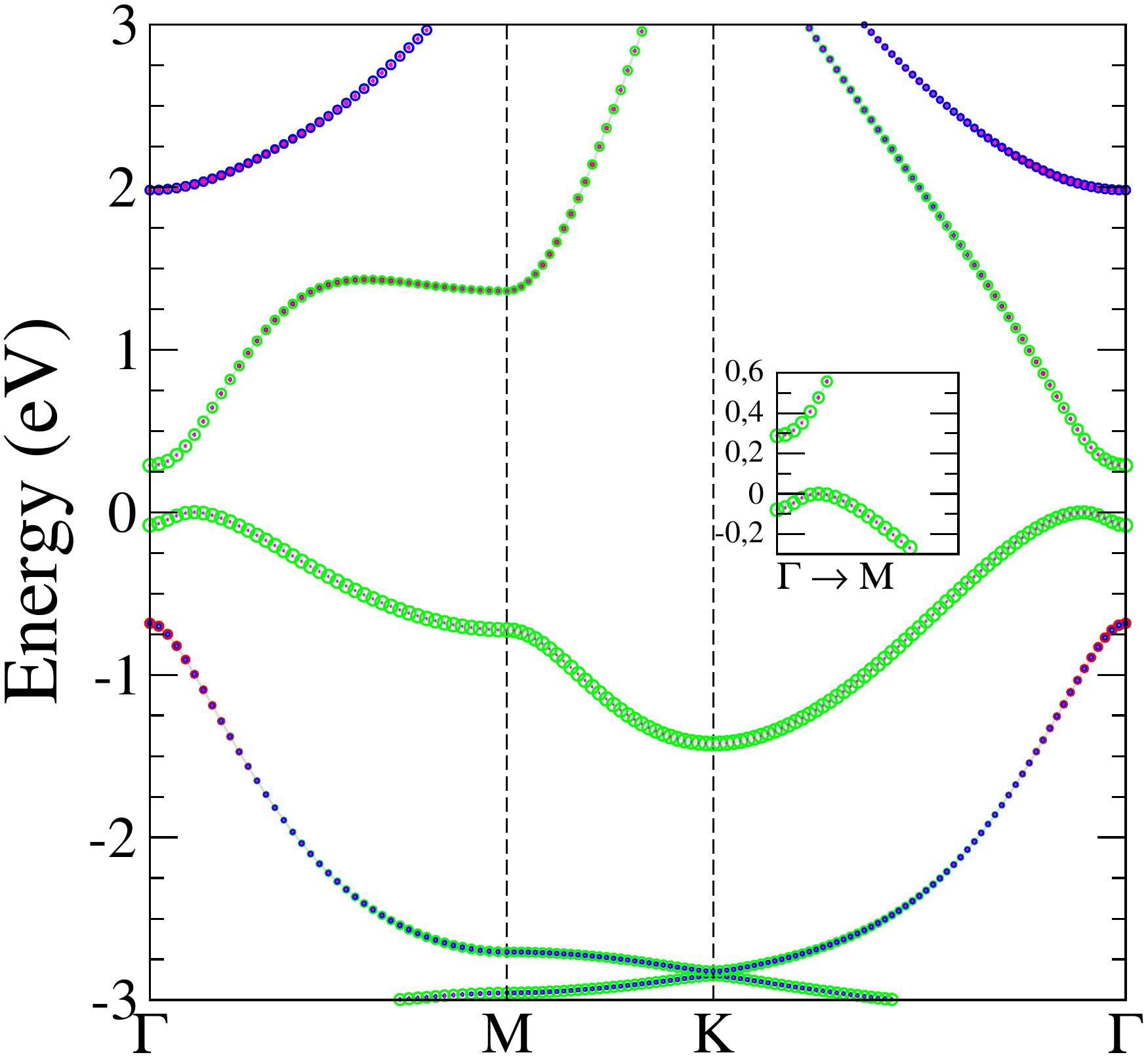}}\qquad
	\subfigure[{} stanene (rectangular)]{\includegraphics[scale=0.31]{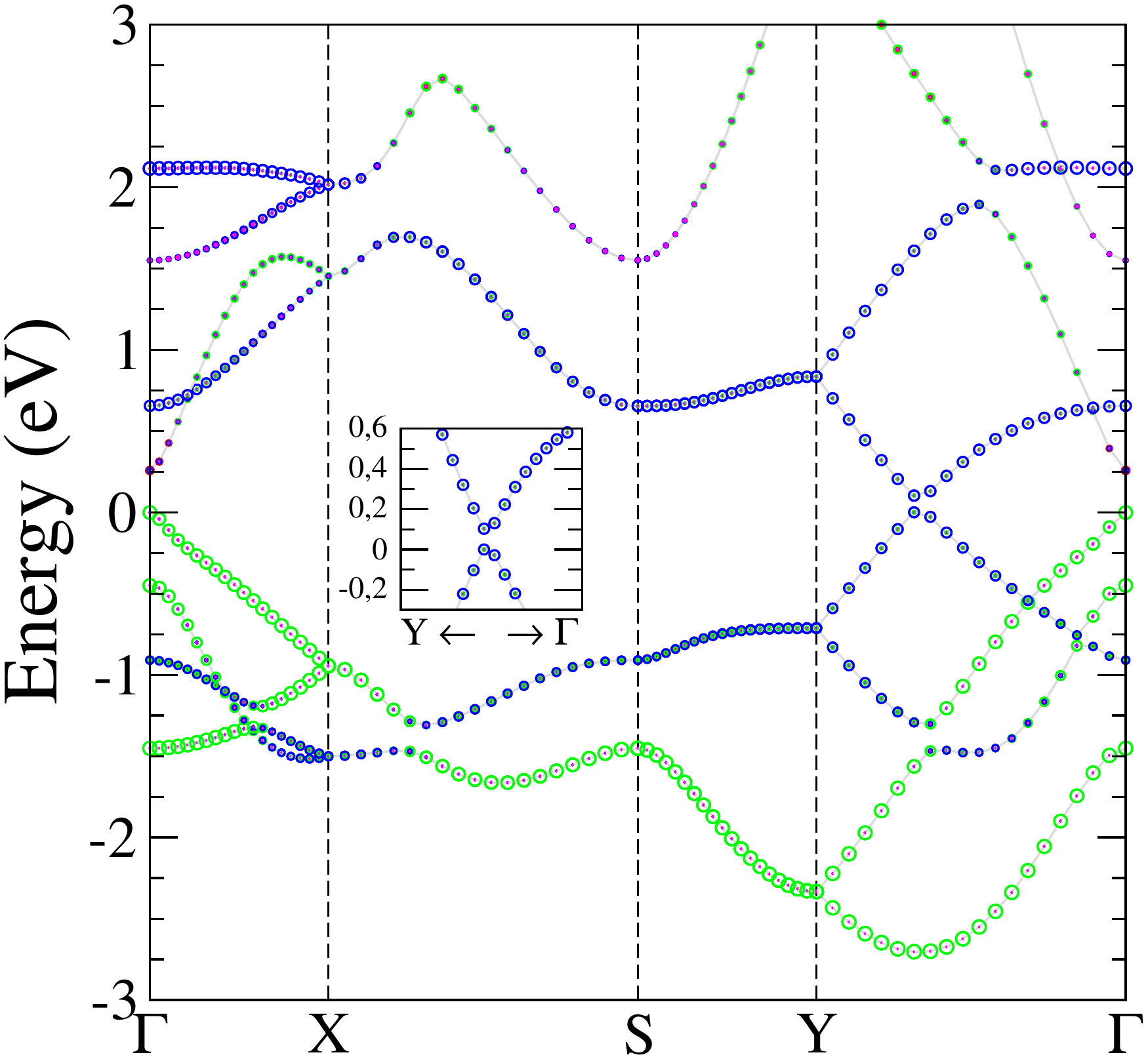}}

	\caption{\label{band-hex2} Band structures for (a) stanene, (b) fluorostanene (hexagonal cell) and (c) stanene (rectangular cell). Red, green, blue and magenta circles depict relative contributions from $s$, $p_x+p_y$ , $p_z$ and $d$ orbitals, respectively, to the band character. The insets display the gap regions.}
\end{figure}

\begin{figure}[H]
	\centering
	\subfigure[{} 1S-MoS$_2$]{\includegraphics[scale=0.08]{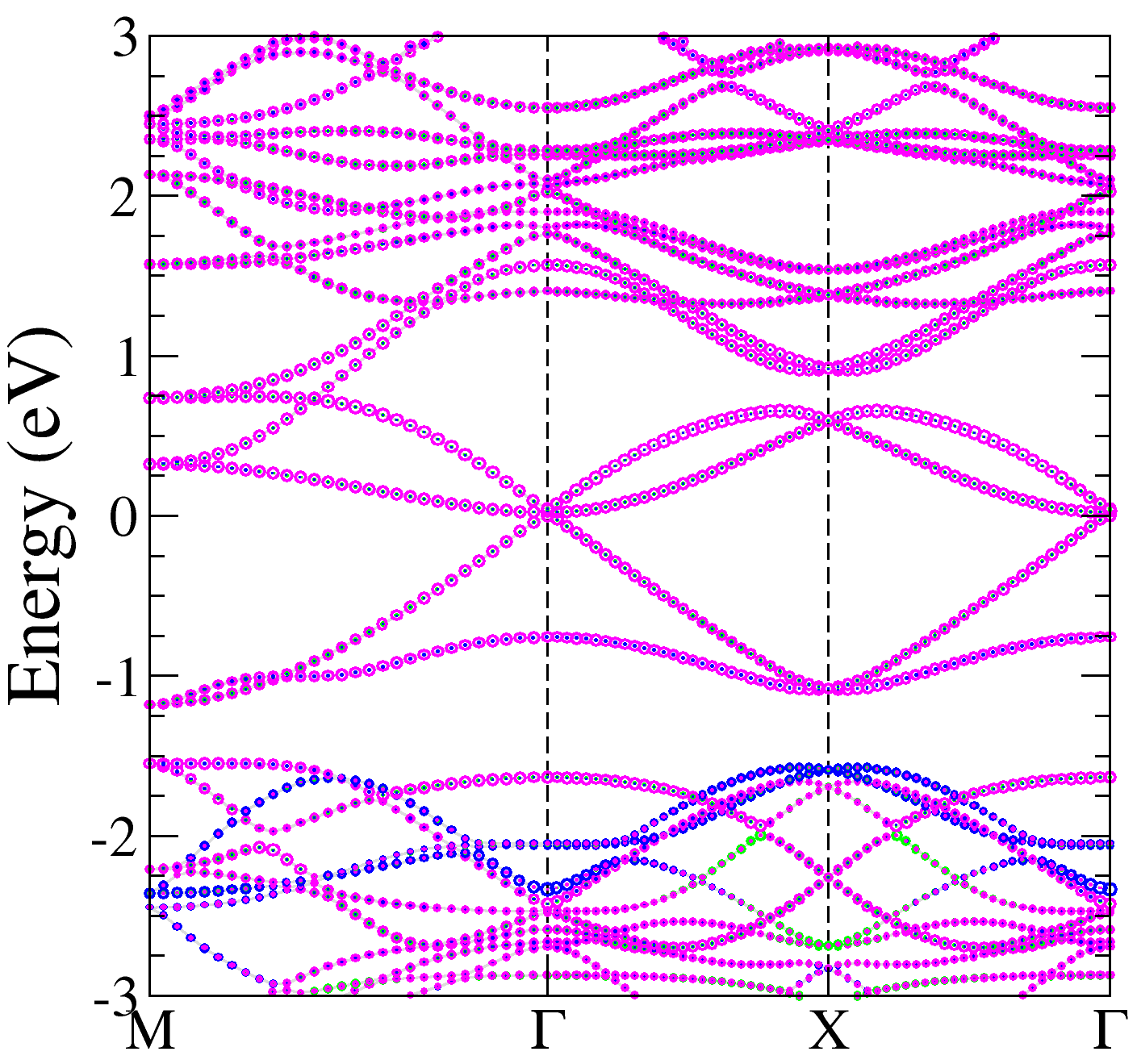}}
	\subfigure[{} 1S-WS$_2$]{\includegraphics[scale=0.08]{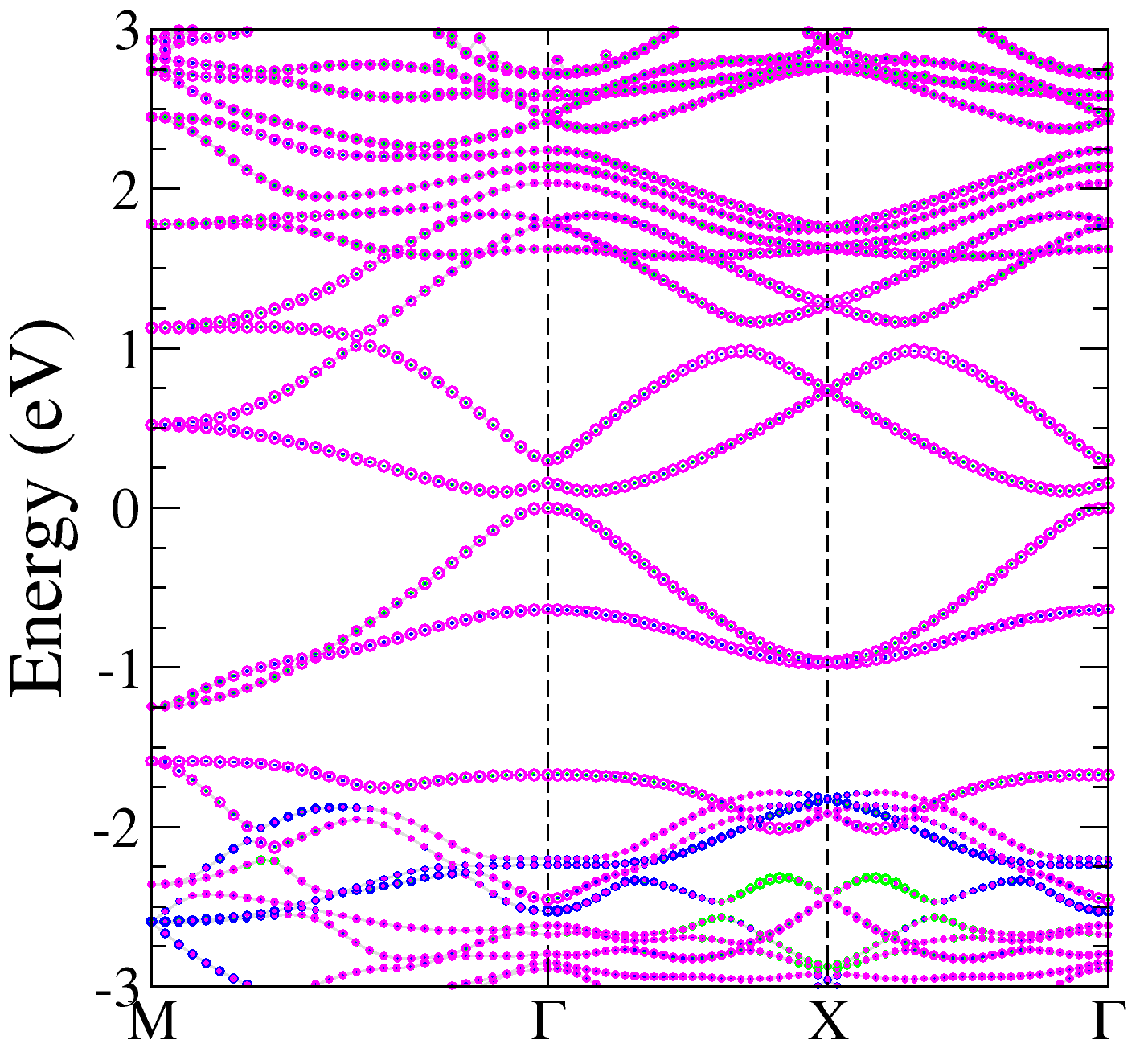}}
	\subfigure[{} 1T'-MoS$_2$]{\includegraphics[scale=0.08]{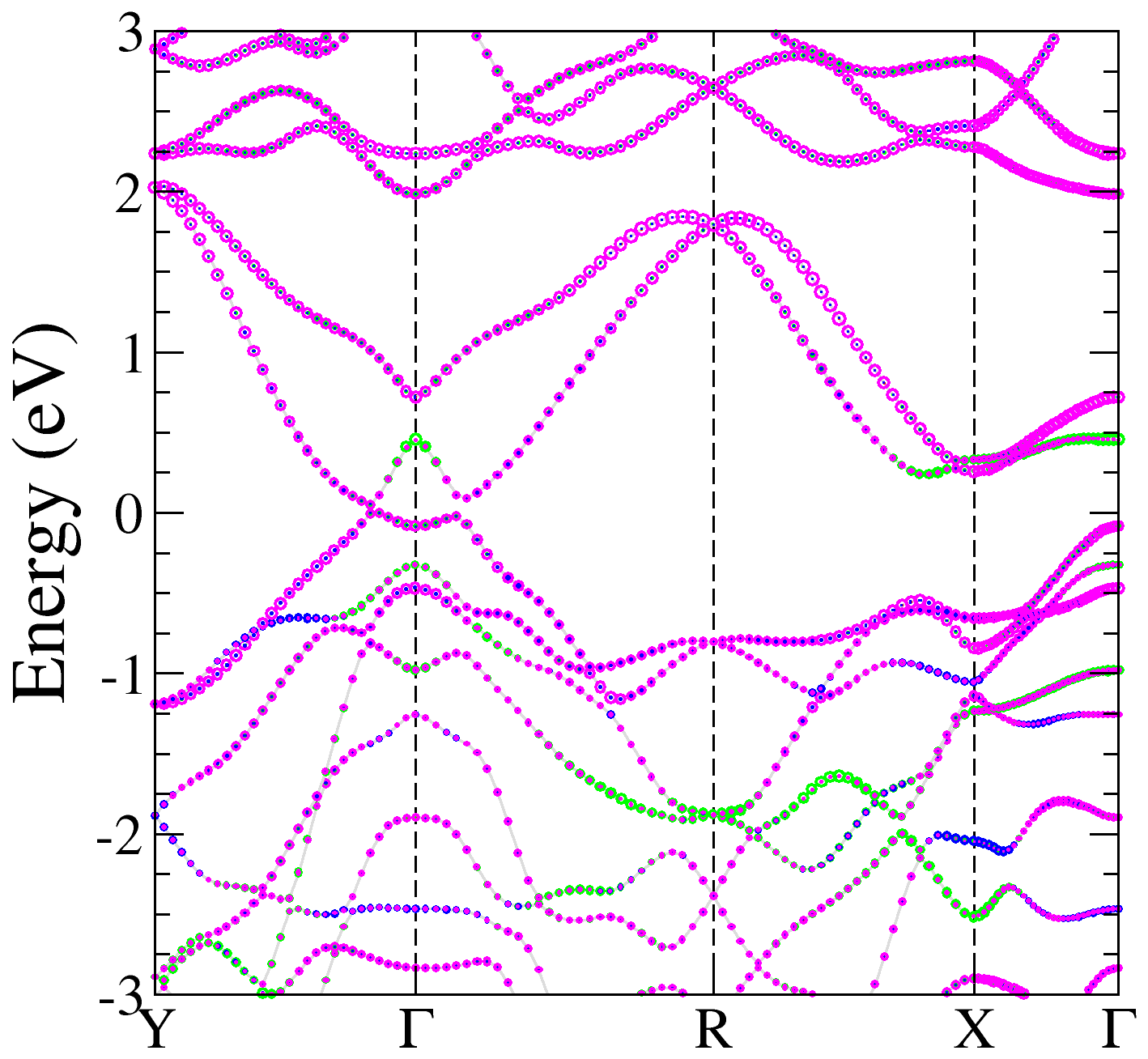}}
	\subfigure[{} 1T'-WS$_2$]{\includegraphics[scale=0.08]{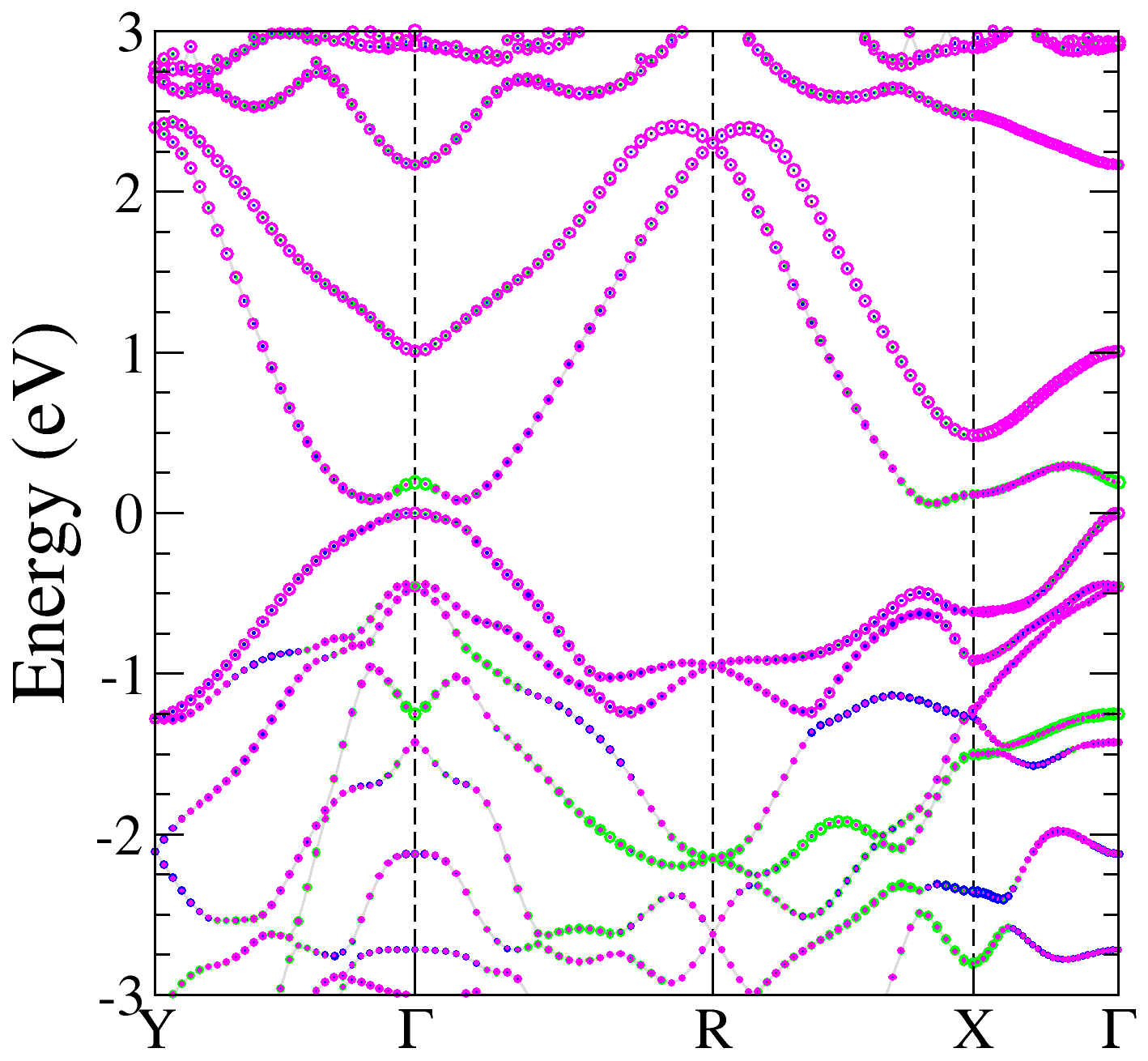}}
	\caption{\label{band-rec} Band structures for (a) 1S-MoS$_2$, (b) 1S-WS$_2$, (c) 1T'-MoS$_2$ and (d) 1T'-WS$_2$. Red, green, blue and magenta circles depict relative contributions from $s$, $p_x+p_y$ , $p_z$ and $d$ orbitals, respectively, to the band character. }
\end{figure}

\begin{figure}[H]
	\centering
	\subfigure[{} 1S-MoS$_2$]{\includegraphics[scale=0.48]{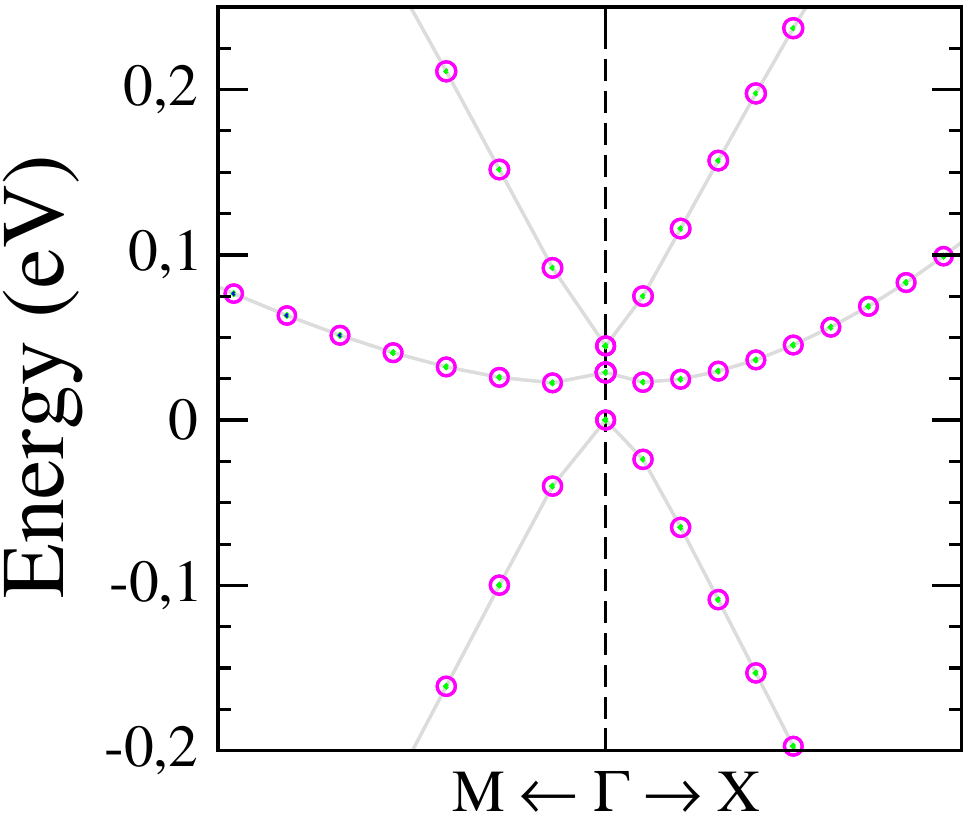}}
	\subfigure[{} 1S-WS$_2$]{\includegraphics[scale=0.48]{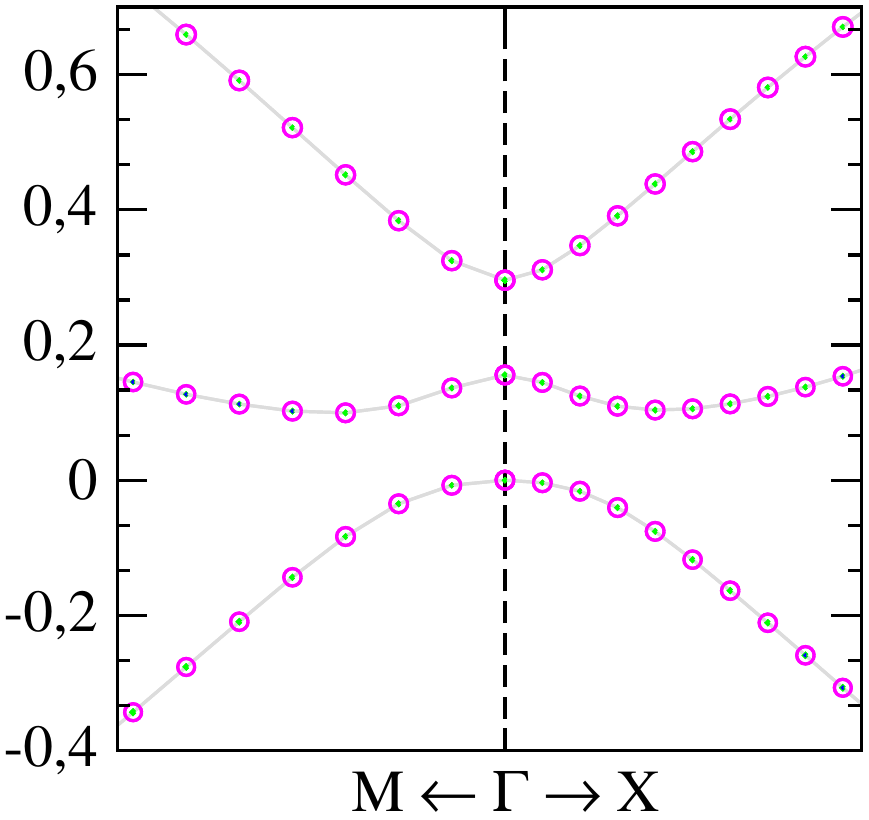}}
	\subfigure[{} 1T'-MoS$_2$]{\includegraphics[scale=0.48]{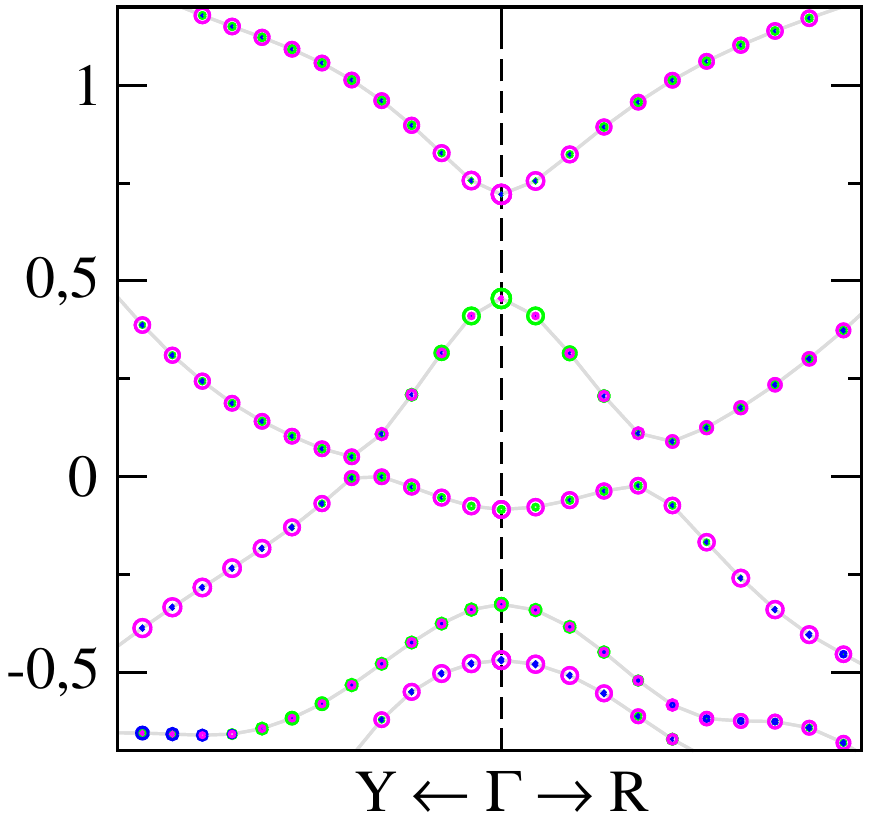}}
	\subfigure[{} 1T'-WS$_2$]{\includegraphics[scale=0.48]{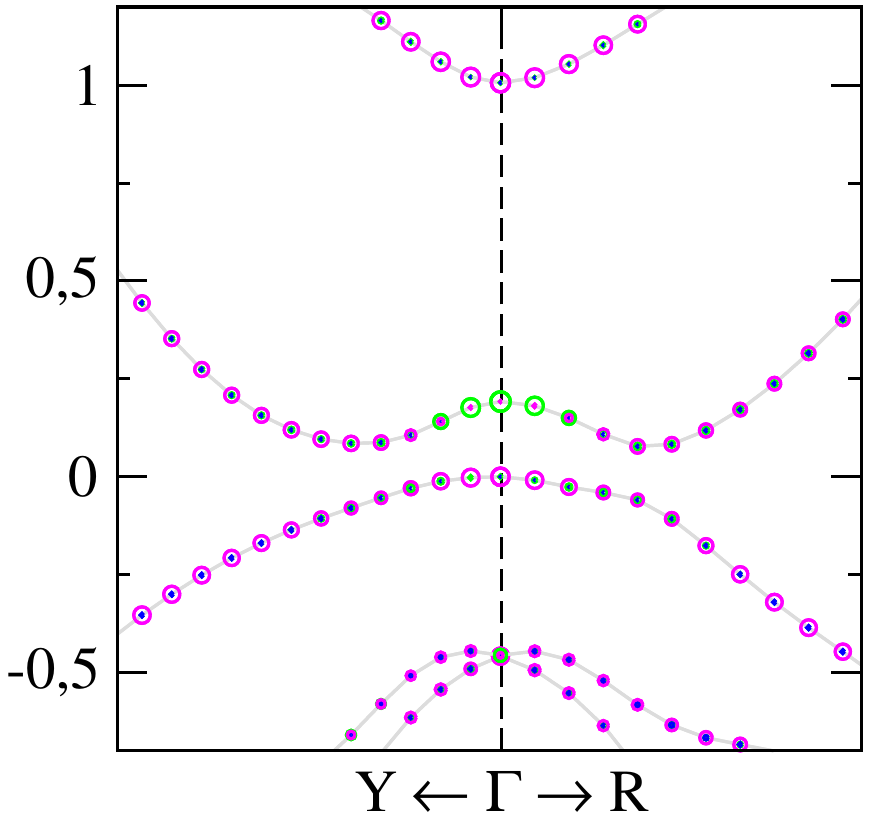}}
	\caption{\label{band-sqr} Small energy window, around the $\Gamma$ point, of the band structures for (a) 1S-MoS$_2$, (b) 1S-WS$_2$, (c) 1T'-MoS$_2$ and (d) 1T'-WS$_2$. Red, green, blue and magenta circles depict relative contributions from $s$, $p_x+p_y$ , $p_z$ and $d$ orbitals, respectively, to the band character. }
\end{figure}

\begin{figure}[ht]
	\centering
	\subfigure[{} F$_z$ = $0$]{\includegraphics[scale=0.33]{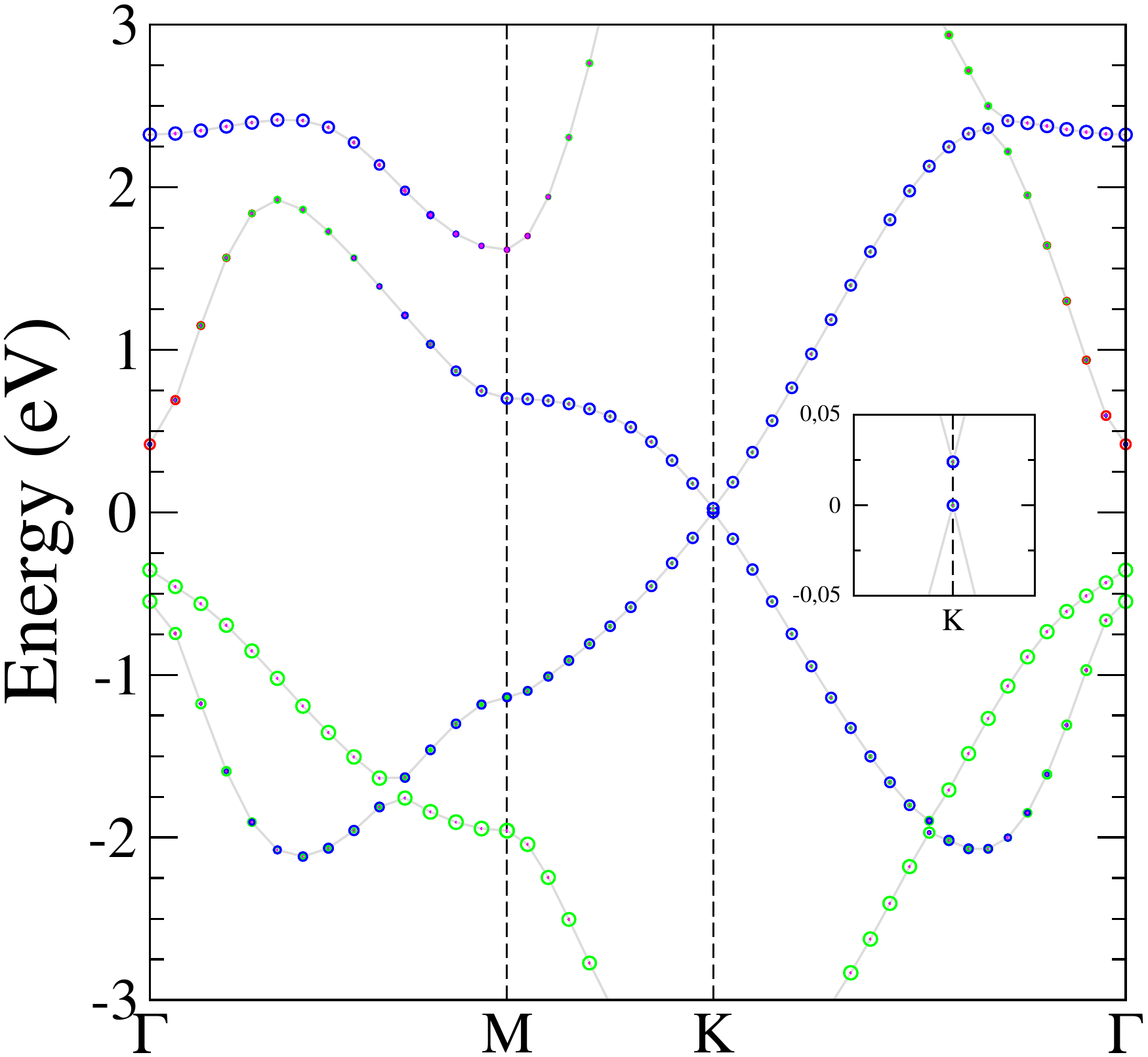}}
	\subfigure[{} F$_z$ = $2\times 10^6$ V/cm]{\includegraphics[scale=0.33]{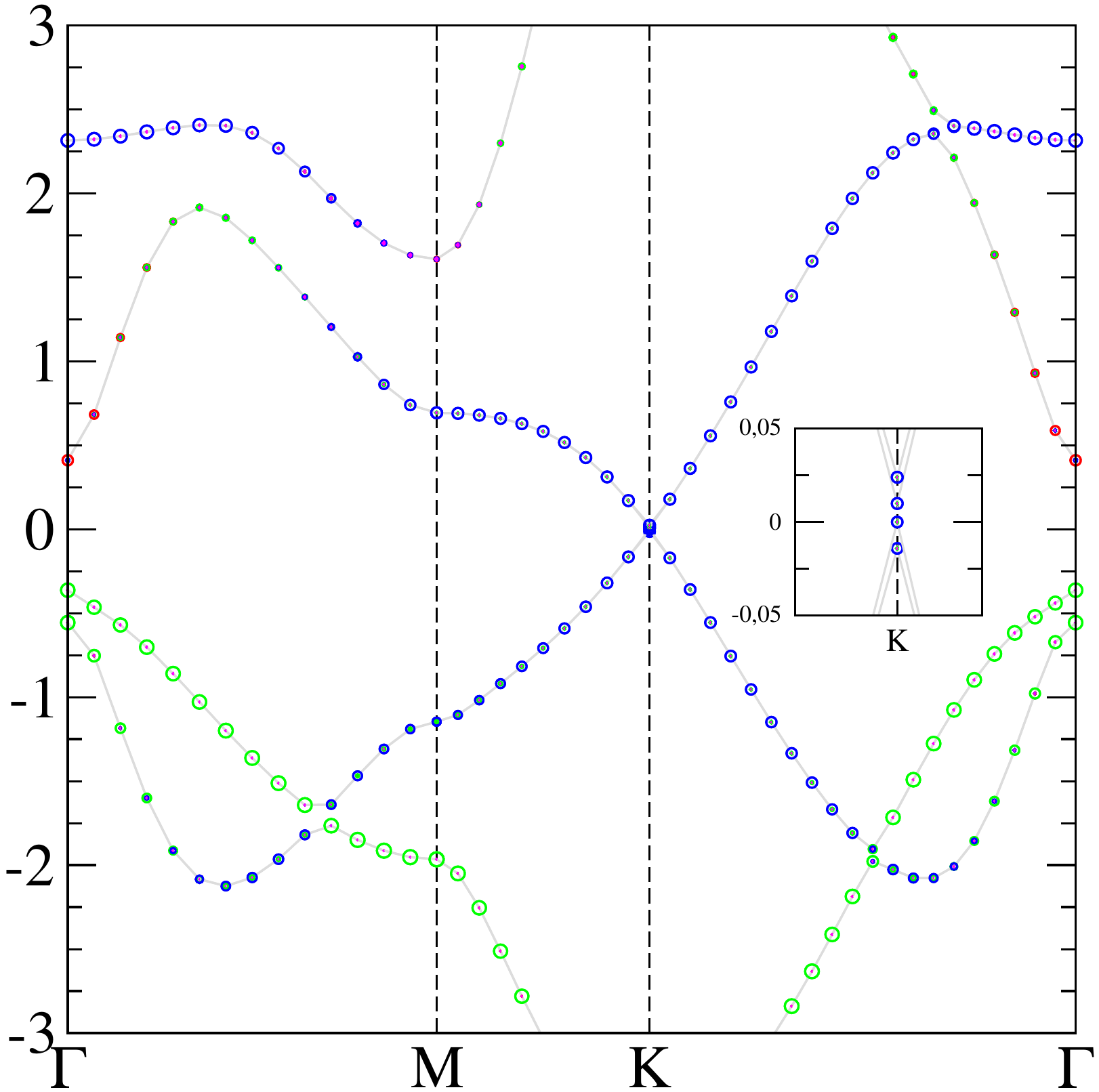}}	
	\subfigure[{} F$_z$ = $5\times 10^6$ V/cm]{\includegraphics[scale=0.33]{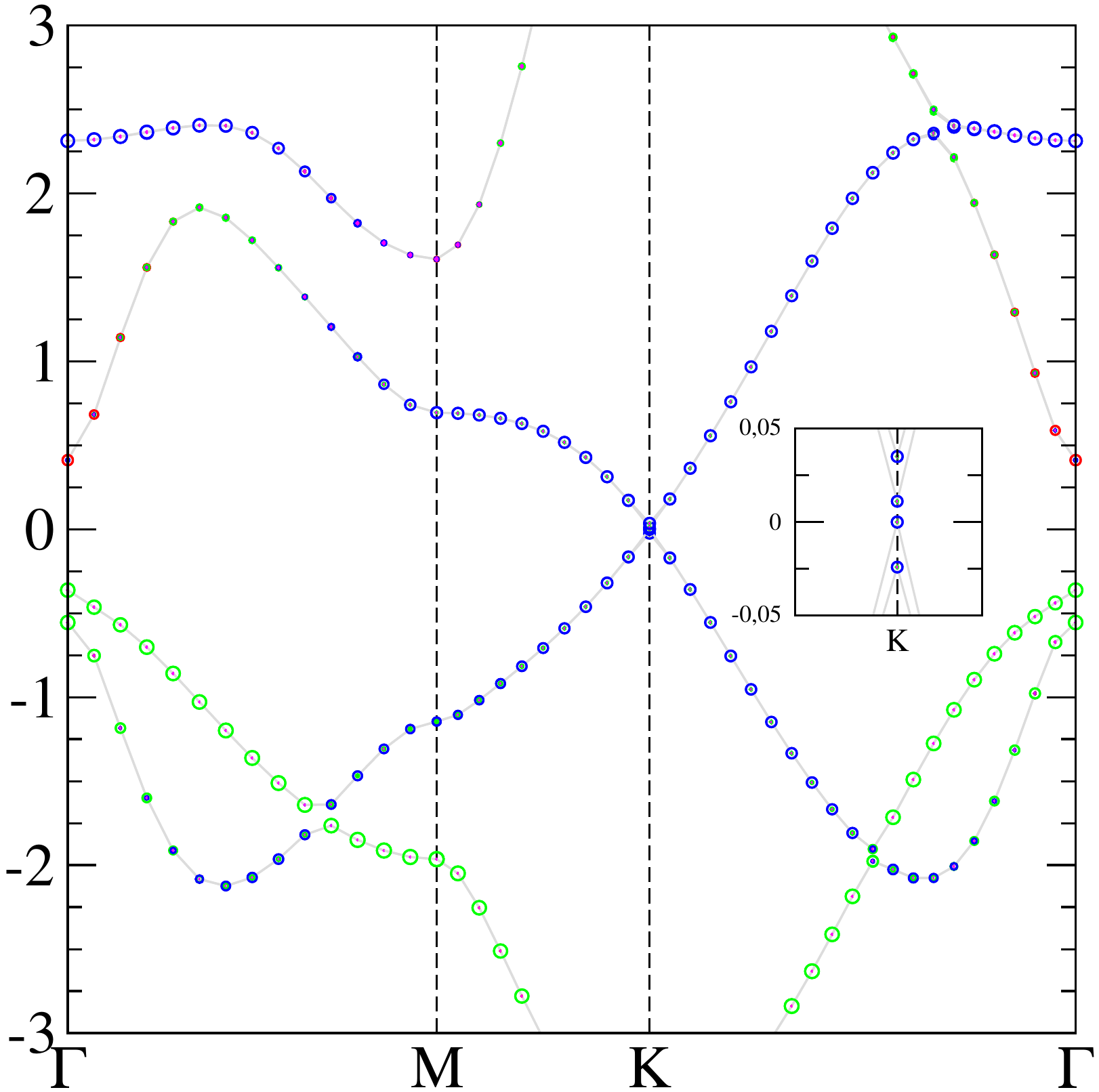}}
	
	\caption{\label{efield} Band structures for (a) germanene, (b) germanene under influence of external electric field E = $2\times 10^6$ V/cm and (c) under E = $5\times 10^6$ V/c. Red, green, blue and magenta circles depict relative contributions from $s$, $p_x+p_y$ , $p_z$ and $d$ orbitals, respectively, to the band character. The insets display the gap regions.}
\end{figure}

\begin{figure}[ht]
	\centering
	\subfigure[{} strain: 0 \%]{\includegraphics[scale=0.24]{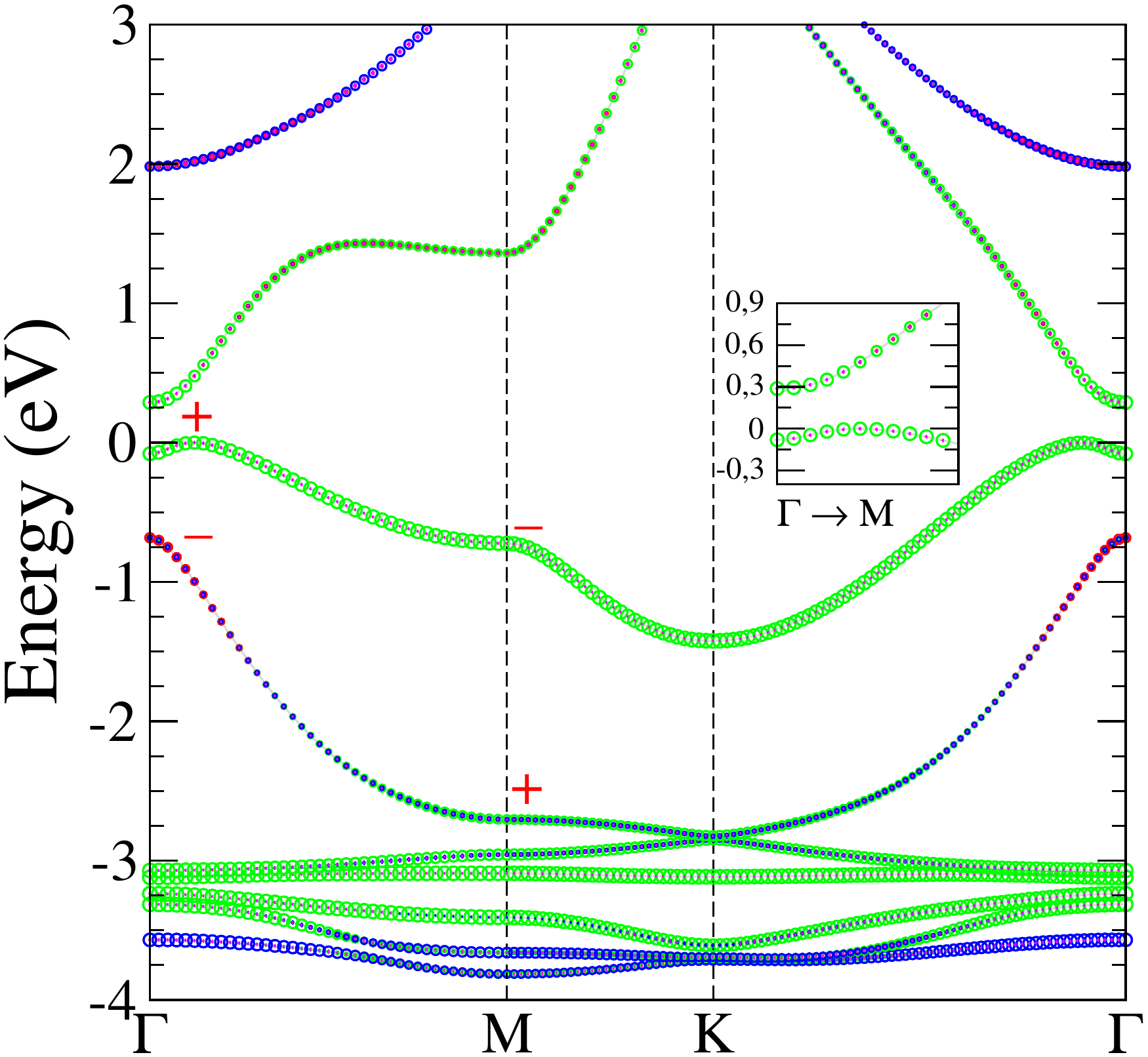}}
	\subfigure[{} strain: 5 \%]{\includegraphics[scale=0.24]{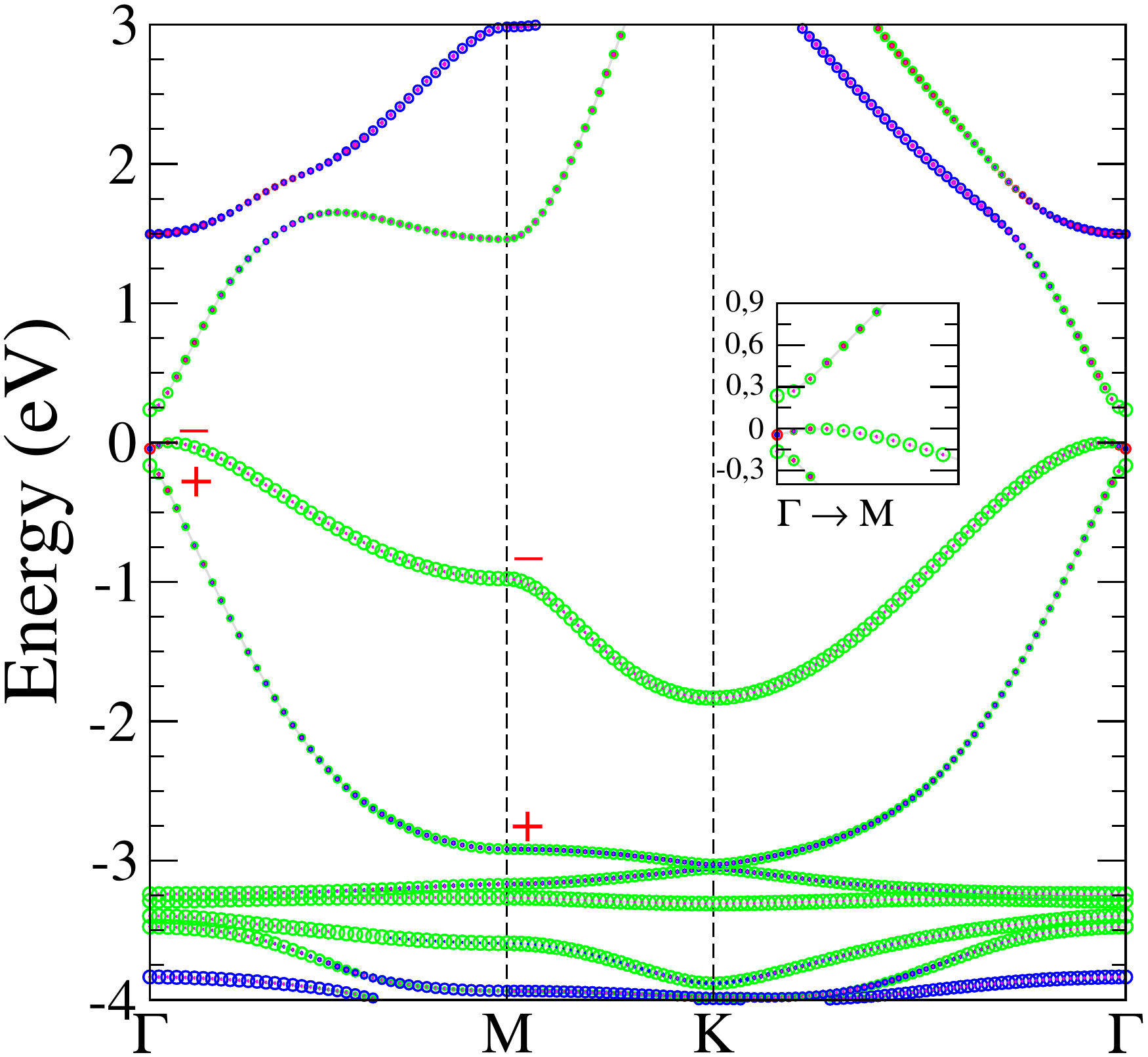}}	
	\subfigure[{} strain: 9 \%]{\includegraphics[scale=0.24]{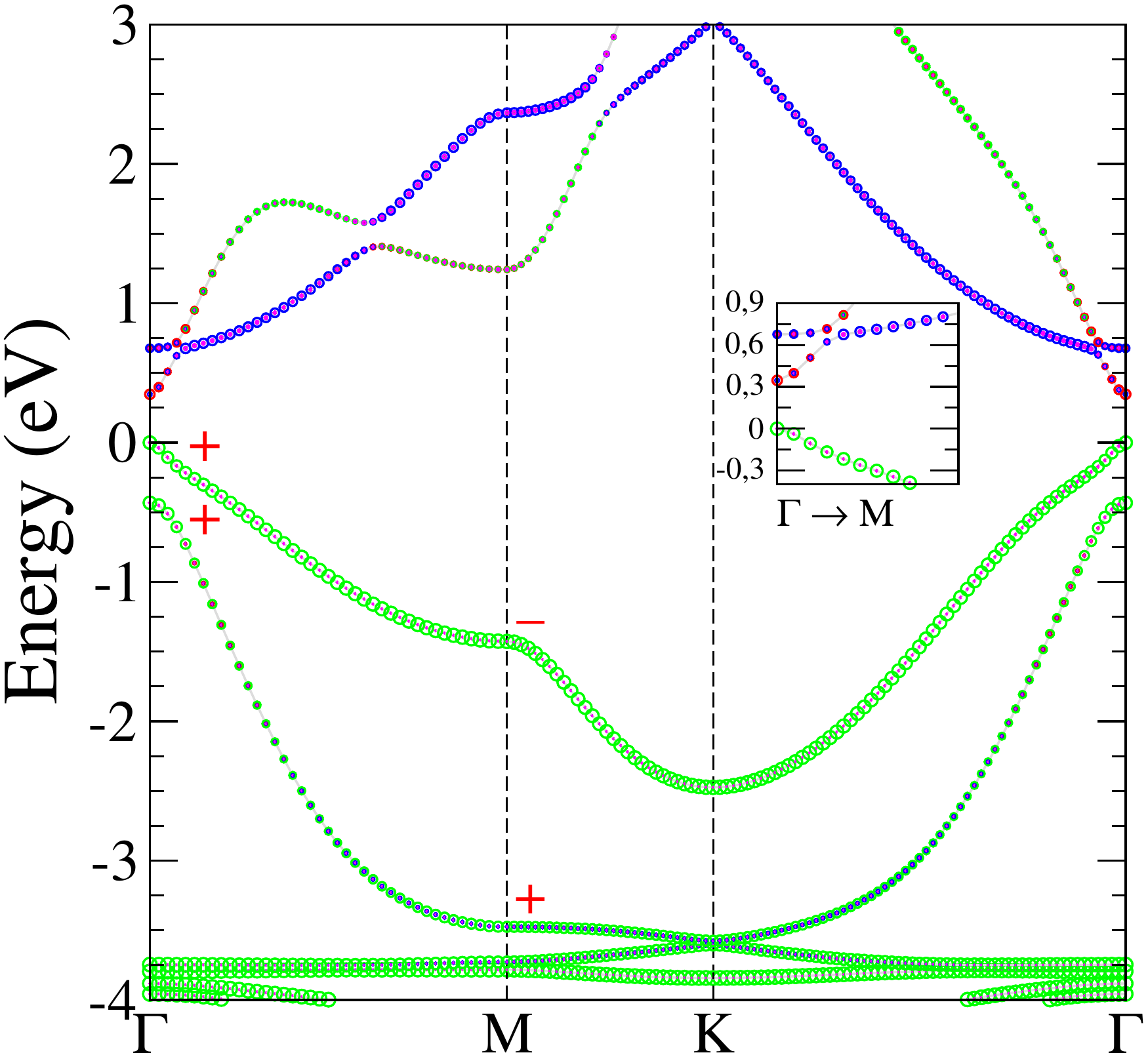}}
	\subfigure[{} strain: 13 \%]{\includegraphics[scale=0.24]{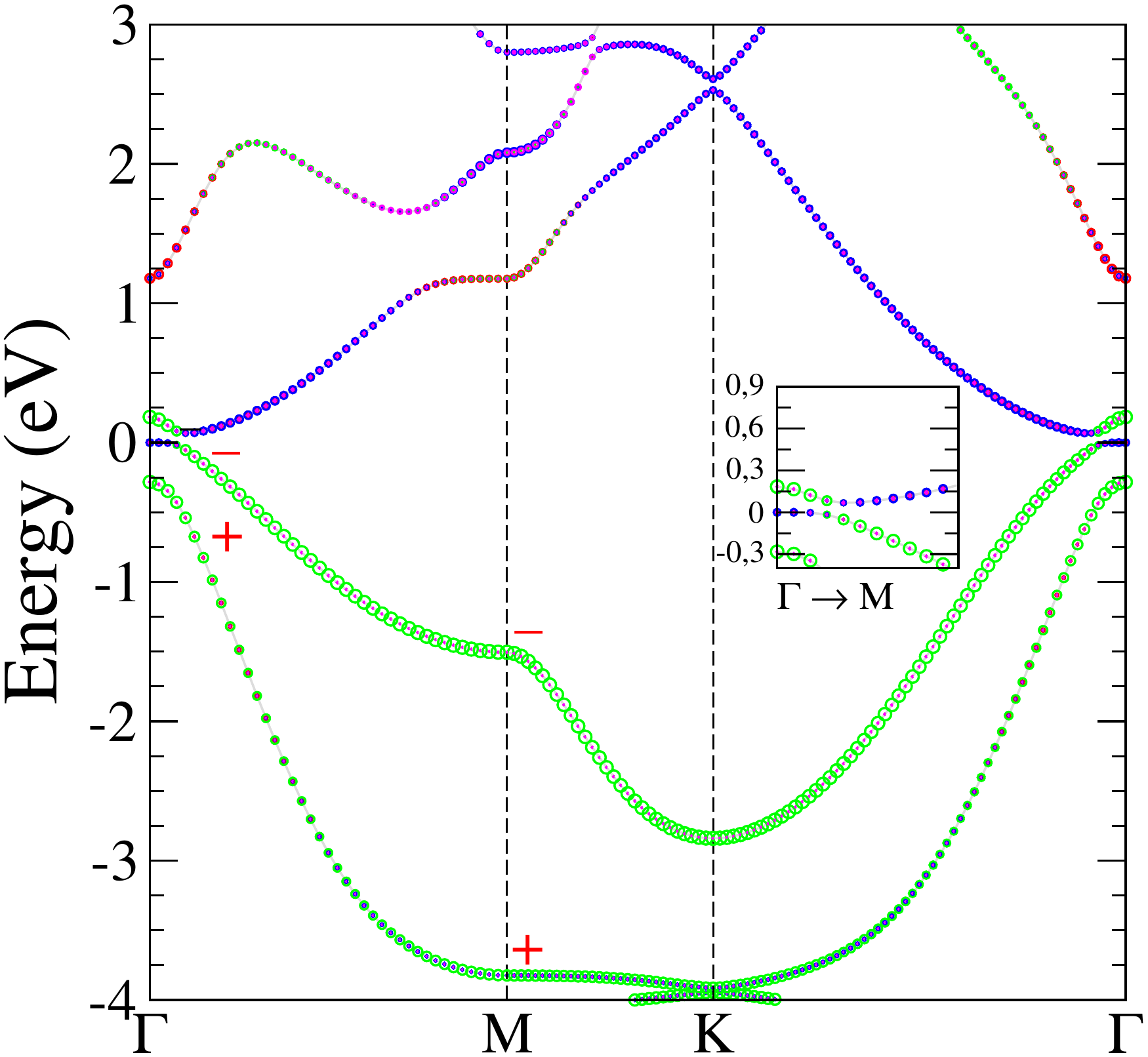}}
		
	\caption{\label{strain} Band structures for (a) fluorostanene, (b) fluorostanene under compressive biaxial strain of 5 \% (c) 9 \% and (d) 13 \%. The plus and minus signs indicate the parity of the highest two occupied bands at $\Gamma$ and $M$ points. The product of the parities of the other bands is always $+1$ at $\Gamma$ and $-1$ at $M$, and do not change with compression. Red, green, blue and magenta circles depict relative contributions from $s$, $p_x+p_y$ , $p_z$ and $d$ orbitals, respectively, to the band character. The insets display the gap regions.}
\end{figure}

\section{Topological properties}

\begin{table}[H]
\centering
\footnotesize
	\caption{\label{tab:s2} Band parities for all hexagonal systems under study at time-reversal invariant momenta ($k_{TRIM}$), and the corresponding Z$_2$ invariants, $\nu$. For hexagonal systems is show only band parities for TRIM points (0.0,0.0) and (0.5,0.5), since the other two are equivalents to (0.5,0.5). For the square lattices 1S-MoS$_2$ and 1S-WS$_2$, the TRIM points (0.5,0.0) and (0.0,0.5) are characterized by an additional degeneracy, in contrast to the other TRIM points with only twofold-degenerate bands (Kramers degeneracy) due to the simultaneous presence of time-reversal and inversion symmetry. The product of their parities, $(\delta_i^2)$, are always $+1$ and do not change the band topology. For the rectangular lattices 1T'-MoS$_2$ and 1T'-WS$_2$ the TRIM points (0.0,0.5) and (0.5,0.5) are also characterized by an additional degeneracy due inversion symmetry and do not contribute to the band topology. }
		\begin{tabular}{|c|c|c|}
			\hline
			\multicolumn{3}{|c|}{germanene} \\
\hline			
			$k_{TRIM}$ &         parity $\xi_{2m}$ of occupied bands         & $\delta_i=\prod_{m}\xi_{2m}$ \\
			\hline
			(0.0,0.0)  &                       $+ - + +$                       &              $-$               \\
			(0.5,0.5)  &                       $- + - +$                       &              $+$               \\
			\hline
			           & Z$_2$ invariant $\nu$: $(-1)^{\nu}=\prod_i\delta_i$ &           $\nu=1$            \\
			\hline
		\end{tabular}
	\quad
		\begin{tabular}{|c|c|c|}
			\hline
			\multicolumn{3}{|c|}{GeI}                                                                   \\
			\hline
			$k_{TRIM}$ &         parity $\xi_{2m}$ of occupied bands         & $\delta_i=\prod_{m}\xi_{2m}$ \\
			\hline
			(0.0,0.0)  &                $+ - + - + + + - - - +$                &              $-$               \\
			(0.5,0.5)  &                $- + - + + - - + - + -$                &              $+$               \\
			\hline
			           & Z$_2$ invariant $\nu$: $(-1)^{\nu}=\prod_i\delta_i$ &           $\nu=1$            \\
			\hline
		\end{tabular}

\vspace{12pt}	
	
		\begin{tabular}{|c|c|c|}
	\hline
	\multicolumn{3}{|c|}{stanene} \\
	\hline			
	$k_{TRIM}$ &         parity $\xi_{2m}$ of occupied bands         & $\delta_i=\prod_{m}\xi_{2m}$ \\
	\hline
	(0.0,0.0)  &                       $+ - + +$                       &              $-$               \\
	(0.5,0.5)  &                       $- + - +$                       &              $+$               \\
	\hline
	& Z$_2$ invariant $\nu$: $(-1)^{\nu}=\prod_i\delta_i$ &           $\nu=1$            \\
	\hline
\end{tabular}
\quad
\begin{tabular}{|c|c|c|}
	\hline
	\multicolumn{3}{|c|}{fluorostanene}                                                             \\
	\hline
	$k_{TRIM}$ &         parity $\xi_{2m}$ of occupied bands         & $\delta_i=\prod_{m}\xi_{2m}$ \\
	\hline
	(0.0,0.0)  &                $+ - + - + + + - - - +$                &              $-$               \\
	(0.5,0.5)  &                $- + - + + - - + - + -$                &              $+$               \\
	\hline
	           & Z$_2$ invariant $\nu$: $(-1)^{\nu}=\prod_i\delta_i$ &           $\nu=1$            \\
	\hline
\end{tabular}	

\vspace{12pt}	

\begin{tabular}{|c|c|c|}
	\hline
	\multicolumn{3}{|c|}{1S-MoS$_2$} \\
	\hline			
	$k_{TRIM}$ &         parity $\xi_{2m}$ of occupied bands         & $\delta_i=\prod_{m}\xi_{2m}$ \\
	\hline
	(0.0,0.0)  & $+ - - + - - + + + - + - - - - - - + + + + + - + + + - - - + - + + + + -$ &              $-$               \\
	(0.5,0.5)  & $- - + + + + - - + + + + - - - - + + + + - - - - - - - - + + + + - - + +$ &              $+$               \\
	\hline
	& Z$_2$ invariant $\nu$: $(-1)^{\nu}=\prod_i\delta_i$ &           $\nu=1$            \\
	\hline
\end{tabular}
\vspace{12pt}

\begin{tabular}{|c|c|c|}
	\hline
	\multicolumn{3}{|c|}{1S-WS$_2$}                                                             \\
	\hline
	$k_{TRIM}$ &         parity $\xi_{2m}$ of occupied bands         & $\delta_i=\prod_{m}\xi_{2m}$ \\
	\hline
	(0.0,0.0)  & $+ - - + - - + + + + - - - - - - - + + + + + - + - + + - - - + + + + + -$ &              $-$               \\
	(0.5,0.5)  & $- - + + + + - - + + - - + + - - + + + + - - - - - - - - + + + + - - + +$ &              $+$               \\
	\hline
	& Z$_2$ invariant $\nu$: $(-1)^{\nu}=\prod_i\delta_i$ &           $\nu=1$            \\
	\hline
\end{tabular}

\vspace{12pt}	

\begin{tabular}{|c|c|c|}
	\hline
	\multicolumn{3}{|c|}{1T'-MoS$_2$}                                                               \\
	\hline
	$k_{TRIM}$ &         parity $\xi_{2m}$ of occupied bands         & $\delta_i=\prod_{m}\xi_{2m}$ \\
	\hline
	(0.0,0.0)  &        $+ - - + - + + - - + + + - - + + - +$        &             $+$              \\
	(0.5,0.0)  &        $+ - - + - + - - + - + + - - - + + +$        &             $-$              \\
	\hline
	           & Z$_2$ invariant $\nu$: $(-1)^{\nu}=\prod_i\delta_i$ &           $\nu=1$            \\
	\hline
\end{tabular}
\quad
\begin{tabular}{|c|c|c|}
	\hline
	\multicolumn{3}{|c|}{1T'-WS$_2$}                                                             \\
	\hline
	$k_{TRIM}$ &         parity $\xi_{2m}$ of occupied bands         & $\delta_i=\prod_{m}\xi_{2m}$ \\
	\hline
	(0.0,0.0)  & $+ - - + - + + - - + + + - - + + - +$ &              $+$               \\
	(0.5,0.0)  & $+ - - + + - - - + - + + - - + - + +$ &              $-$               \\
	\hline
	& Z$_2$ invariant $\nu$: $(-1)^{\nu}=\prod_i\delta_i$ &           $\nu=1$            \\
	\hline
\end{tabular}
\end{table}

\begin{figure}[H]
	\subfigure[{} 1S-MoS$_2$]{\includegraphics[width=0.5\linewidth]{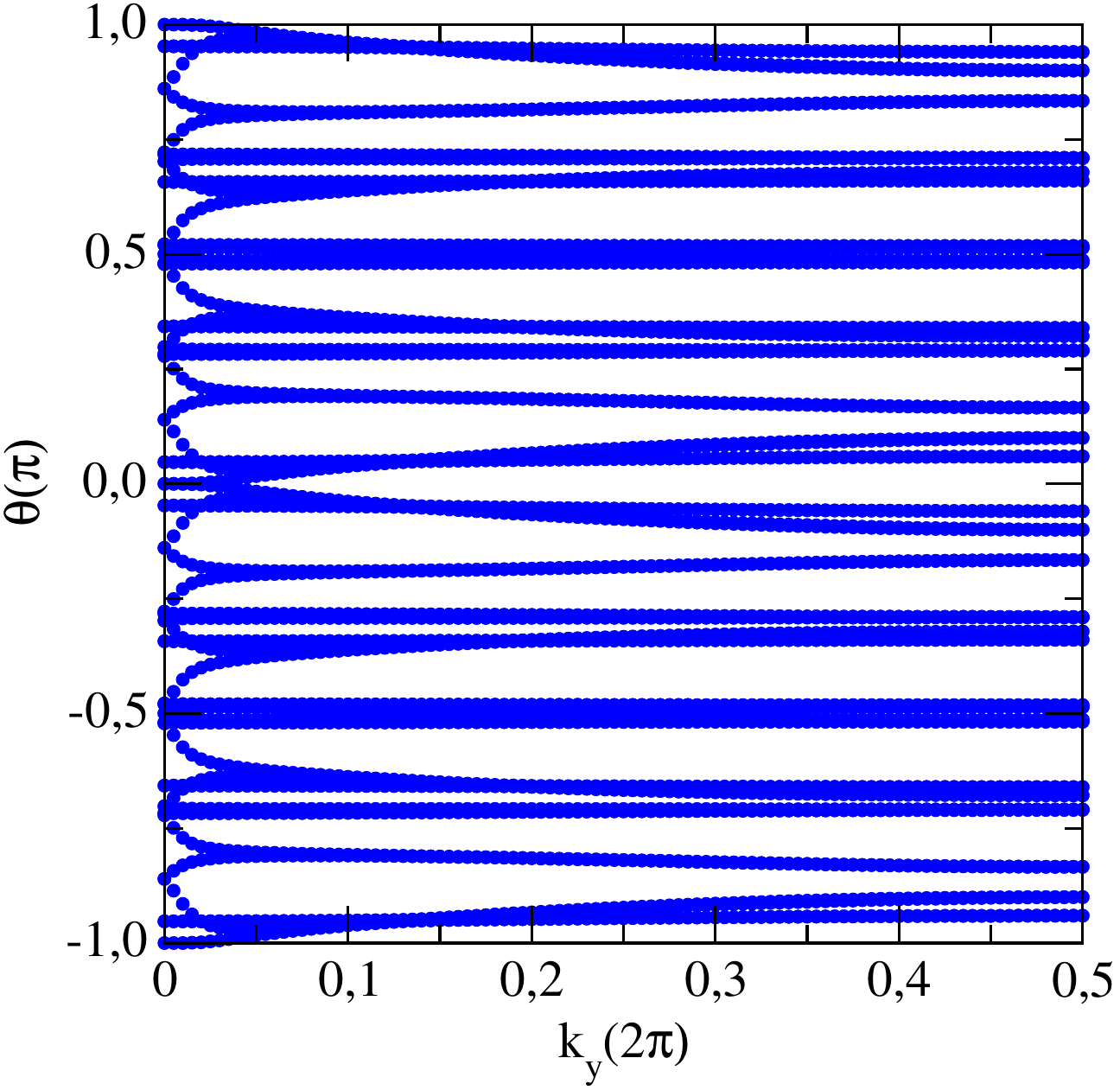}}
	\subfigure[{} 1S-WS$_2$]{\includegraphics[width=0.5\linewidth]{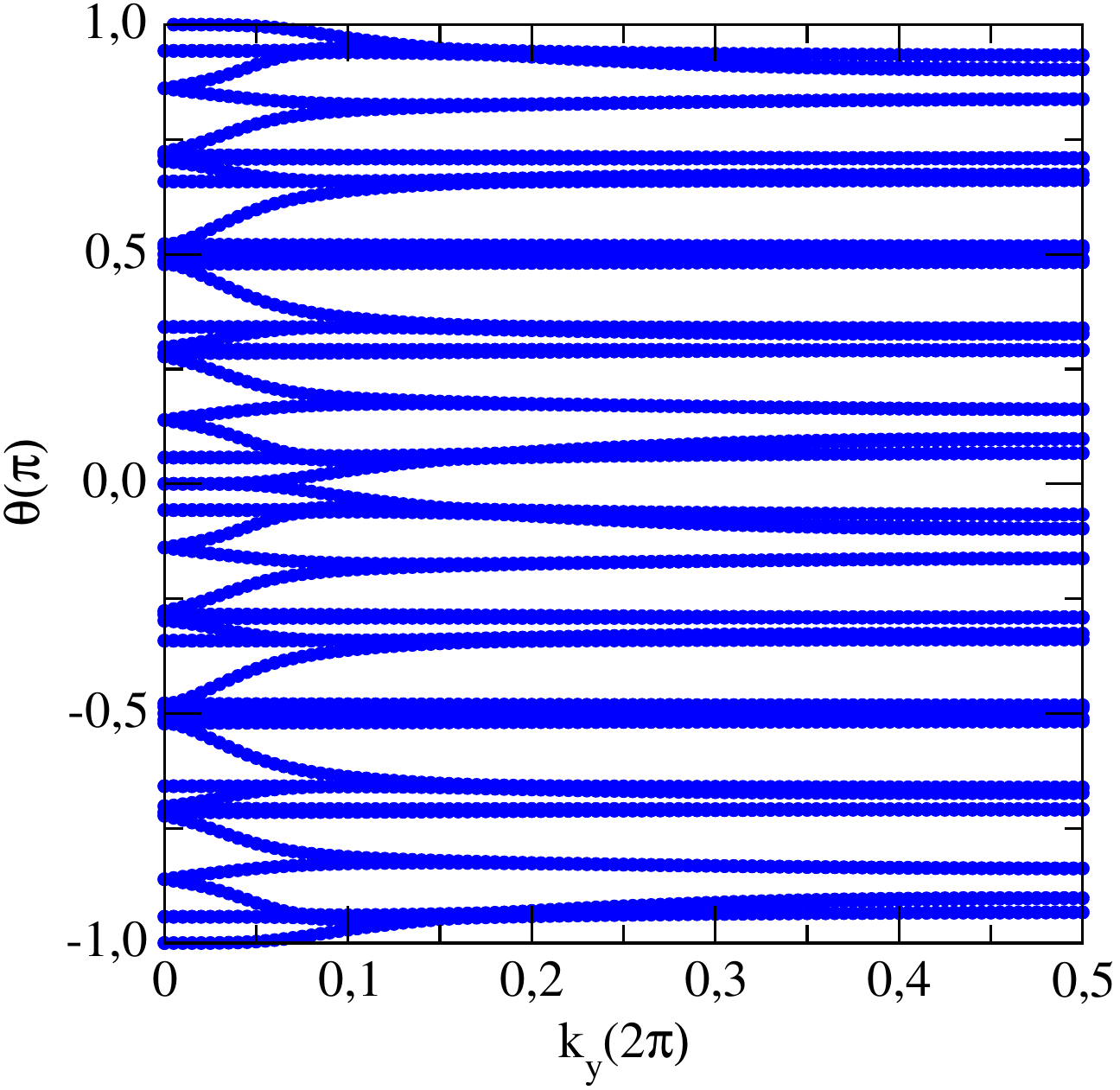}}
	\subfigure[{} 1T'-MoS$_2$]{\includegraphics[width=0.5\linewidth]{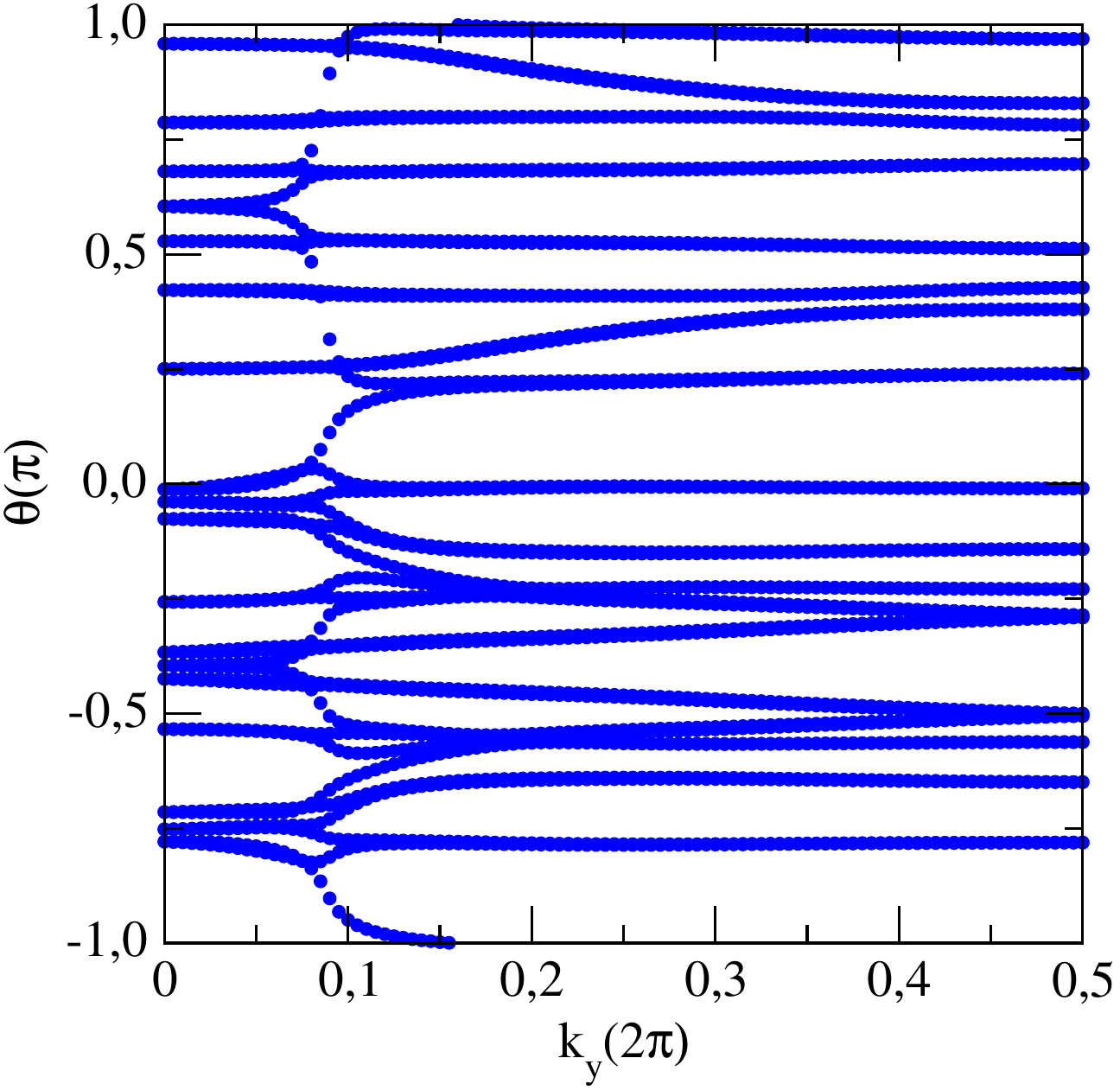}}
	\subfigure[{} 1T'-WS$_2$]{\includegraphics[width=0.5\linewidth]{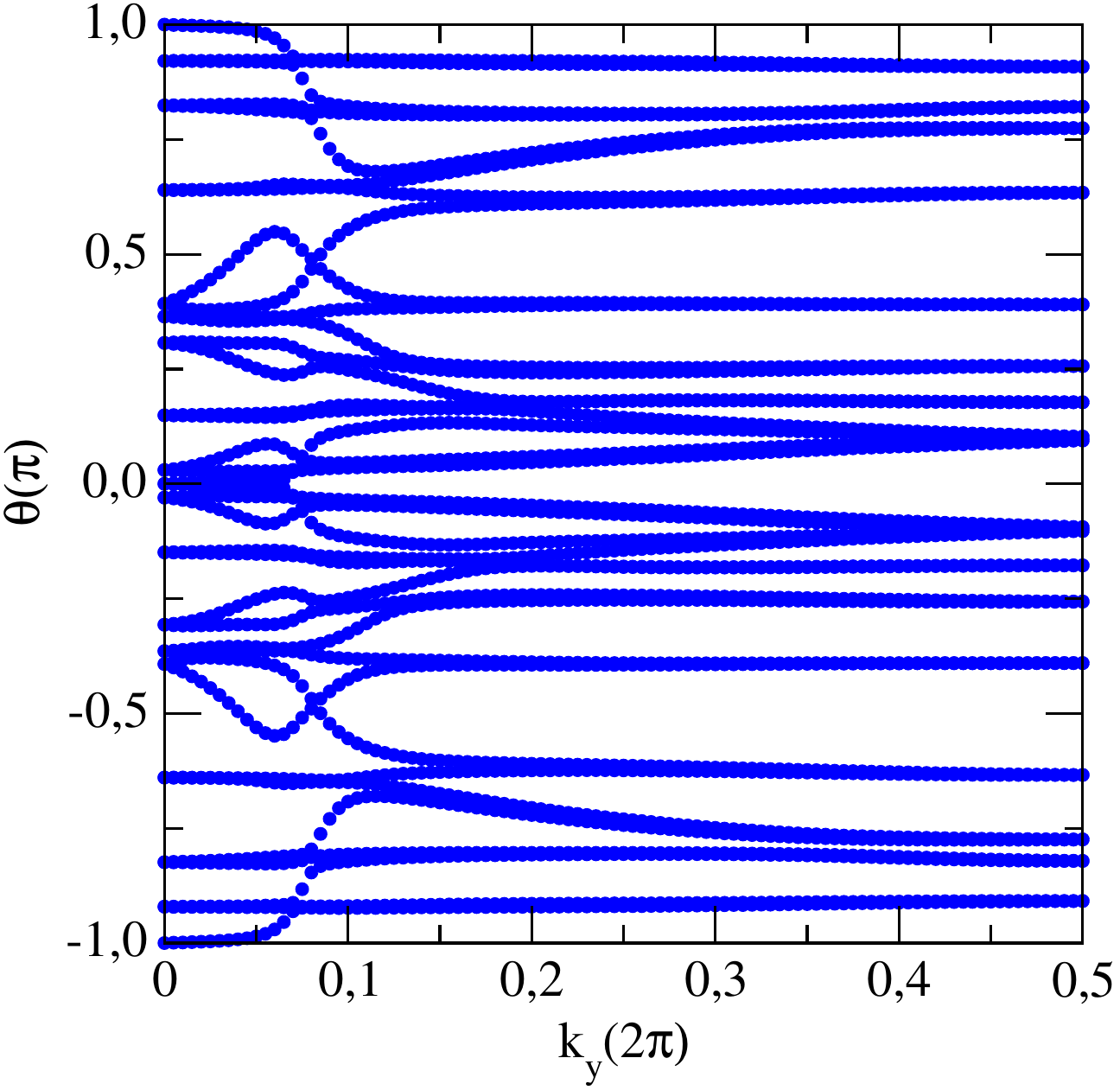}}	
	
	\caption{\label{wcc} Evolution of Wannier charge centers for (a) 1S-MoS$_2$, (b) 1S-WS$_2$, (c) 1T'-MoS$_2$ and (d) 1T'-WS$_2$ along a high-symmetry line between two TRIM points. }
\end{figure}

\section{Static Spin Hall Conductivity}

\begin{figure}[H]
	\centering
	\includegraphics[width=0.7\linewidth]{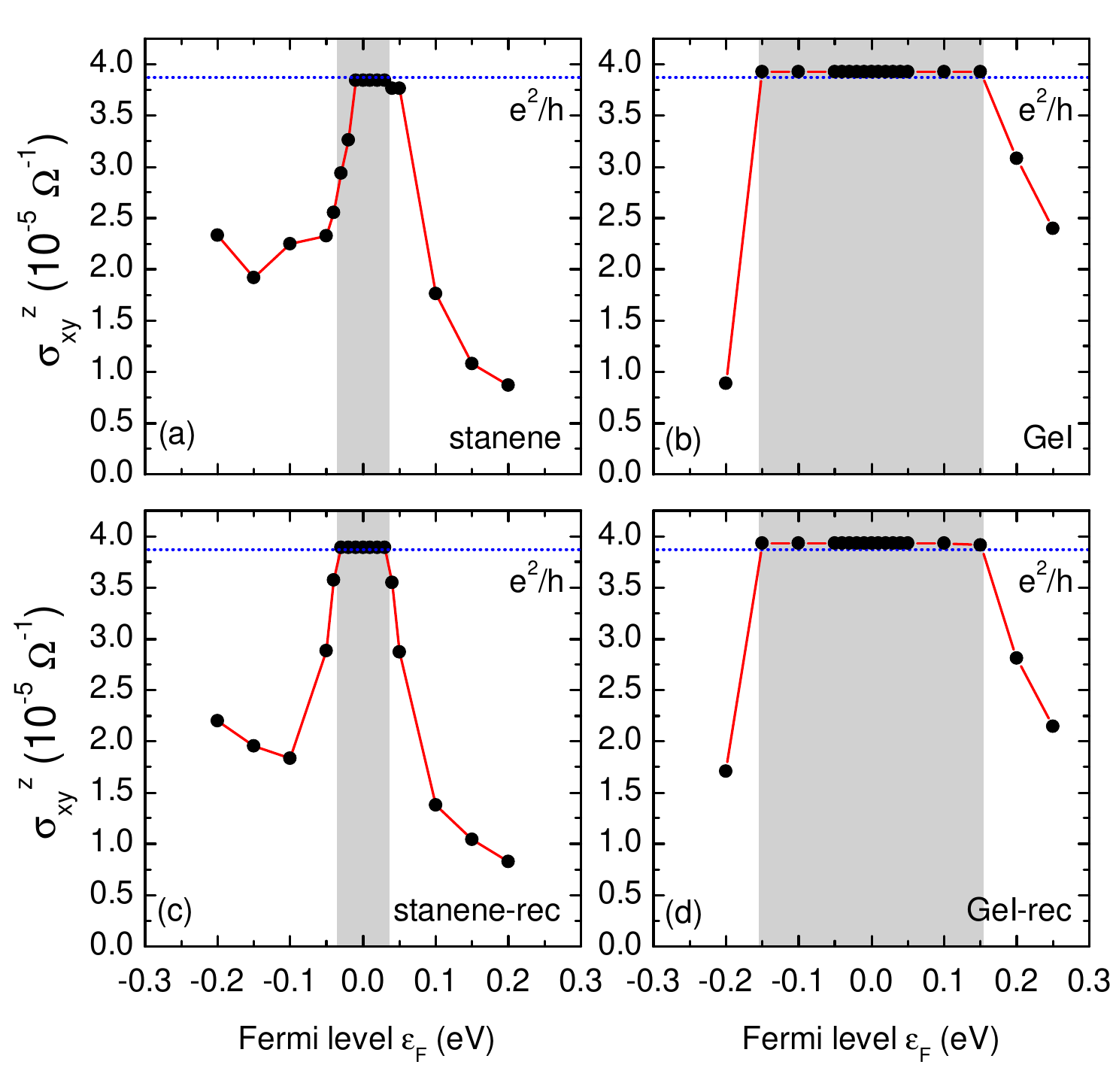}
	\caption{\label{doping} Comparison of the real part of the static spin Hall conductivity $\sigma^z_{xy}$ of stanene (a, c) and GeI (b, d) computed using primitive hexagonal unit cells (a, b) and non-primitive rectangular unit cells (c, d).}
\end{figure}

\section{Frequency-dependent Spin Hall Conductivity}

\begin{figure}[H]
	\centering
	\subfigure[{} germanene]{\includegraphics[scale=0.28]{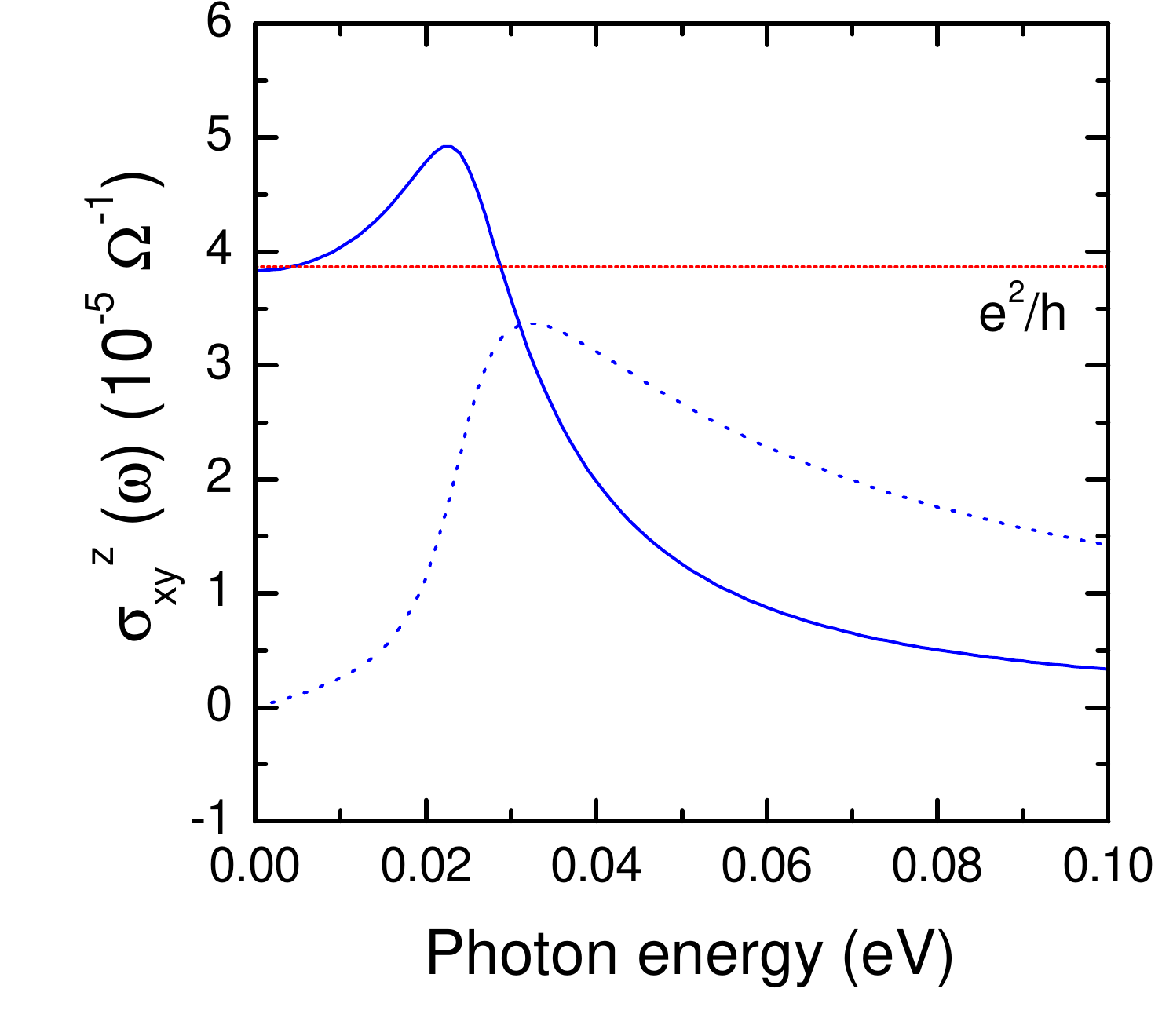}}
	\subfigure[{} GeI]{\includegraphics[scale=0.28]{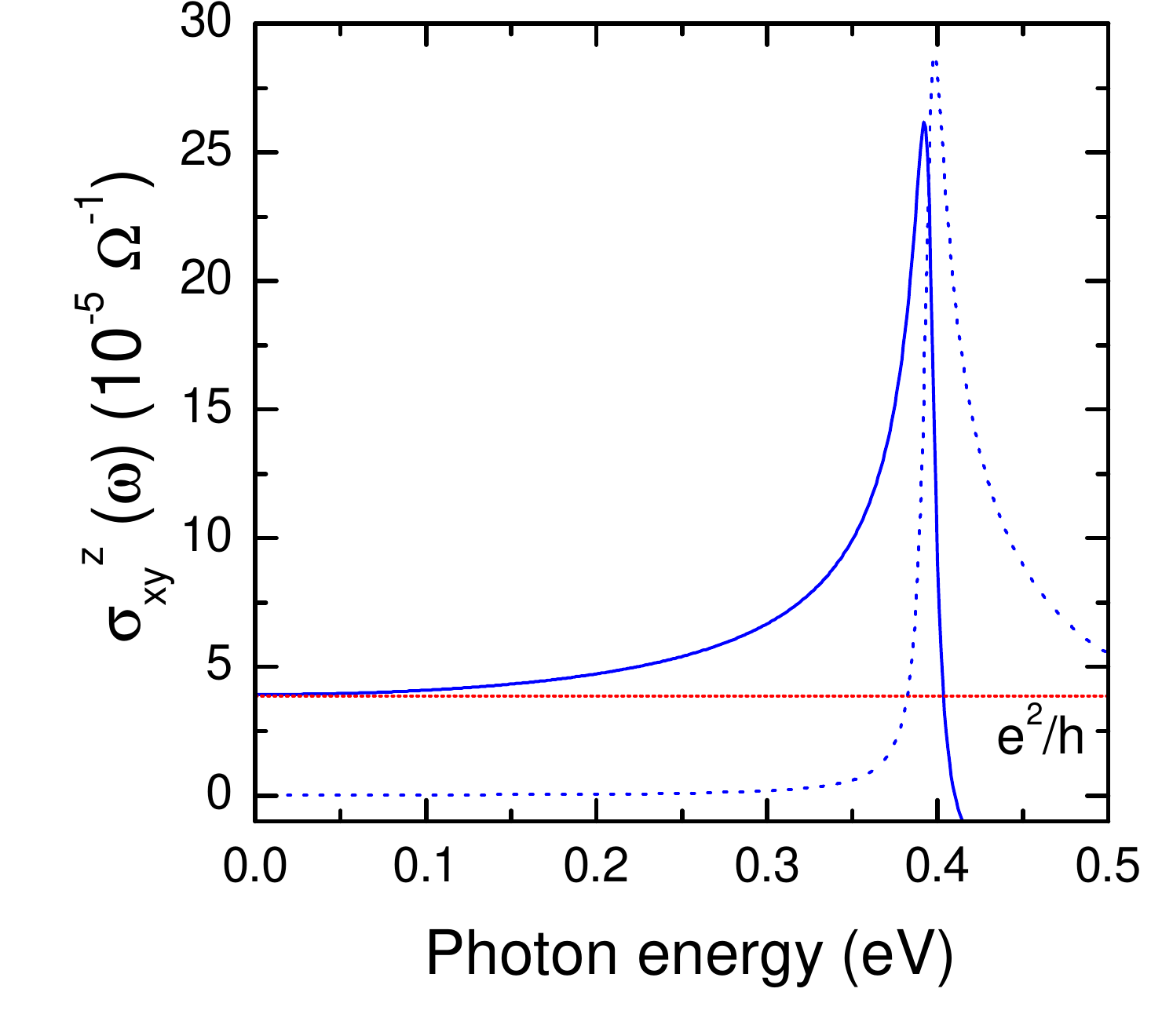}}
	\subfigure[{} stanene]{\includegraphics[scale=0.28]{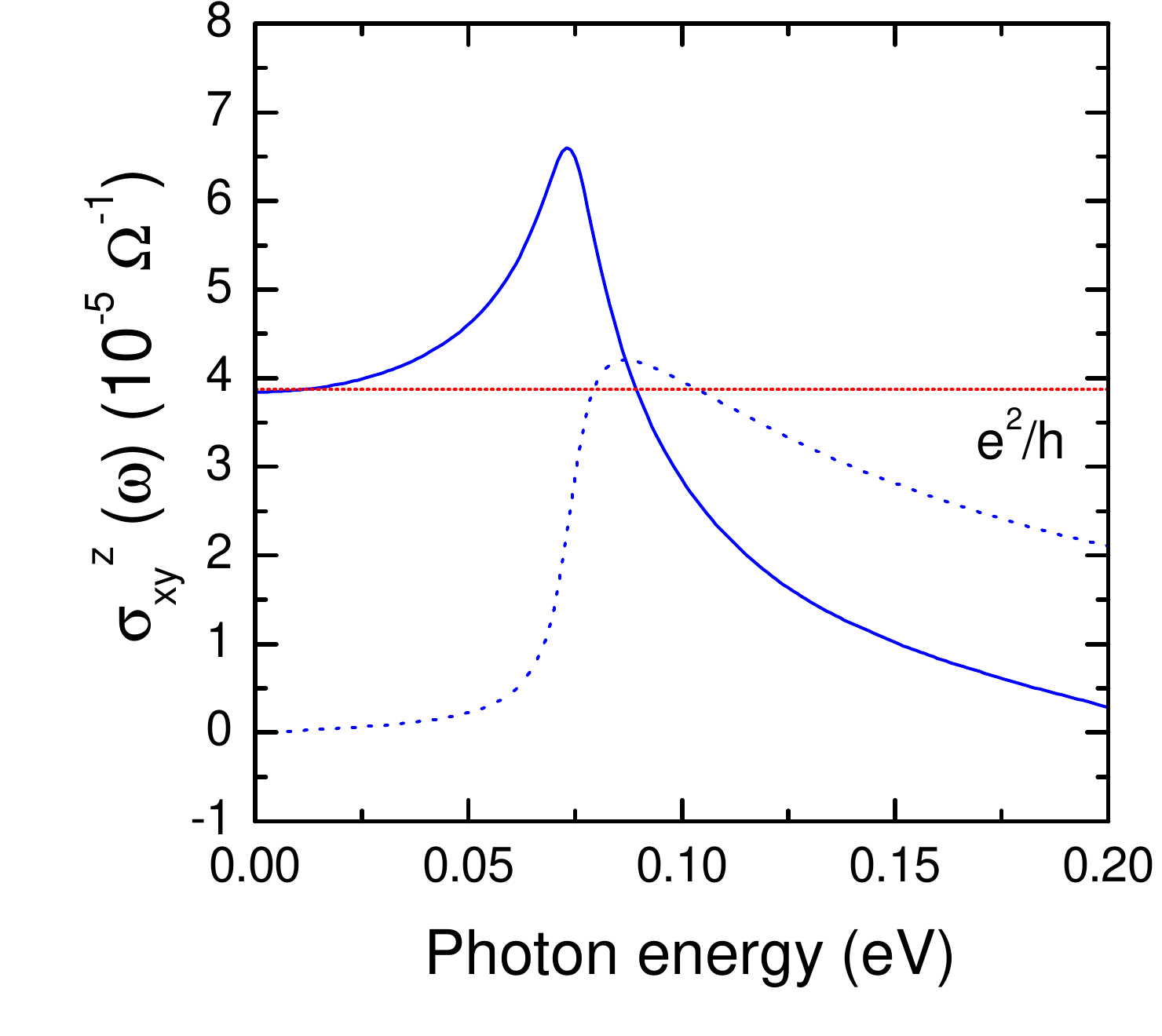}}
	\subfigure[{} fluorostanene]{\includegraphics[scale=0.28]{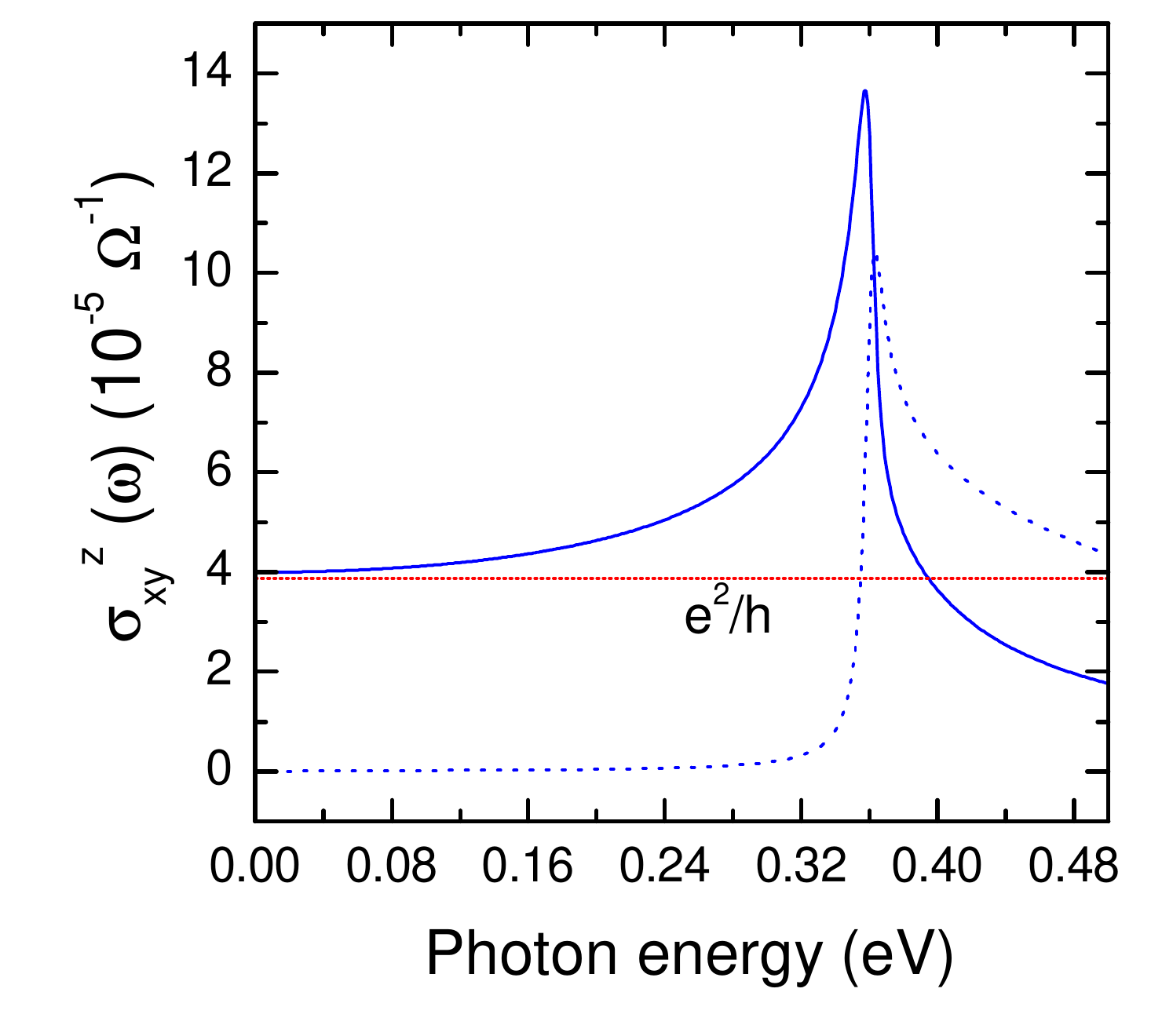}}
	\caption{\label{sigma-freq} Spin Hall conductivity $\sigma^z_{xy}(\omega)$ (real part: solid lines, imaginary part: dashed lines) as a
		function of photon energy for (a) germanene, (b) GeI, (c) stanene and (d) fluorostanene. The sample temperature is set to 0 K. The broadening parameter is fixed at $\eta = 0.005$ eV.}
\end{figure}

\begin{figure}[H]
	\centering
	\subfigure[{} 1S-MoS$_2$]{\includegraphics[scale=0.28]{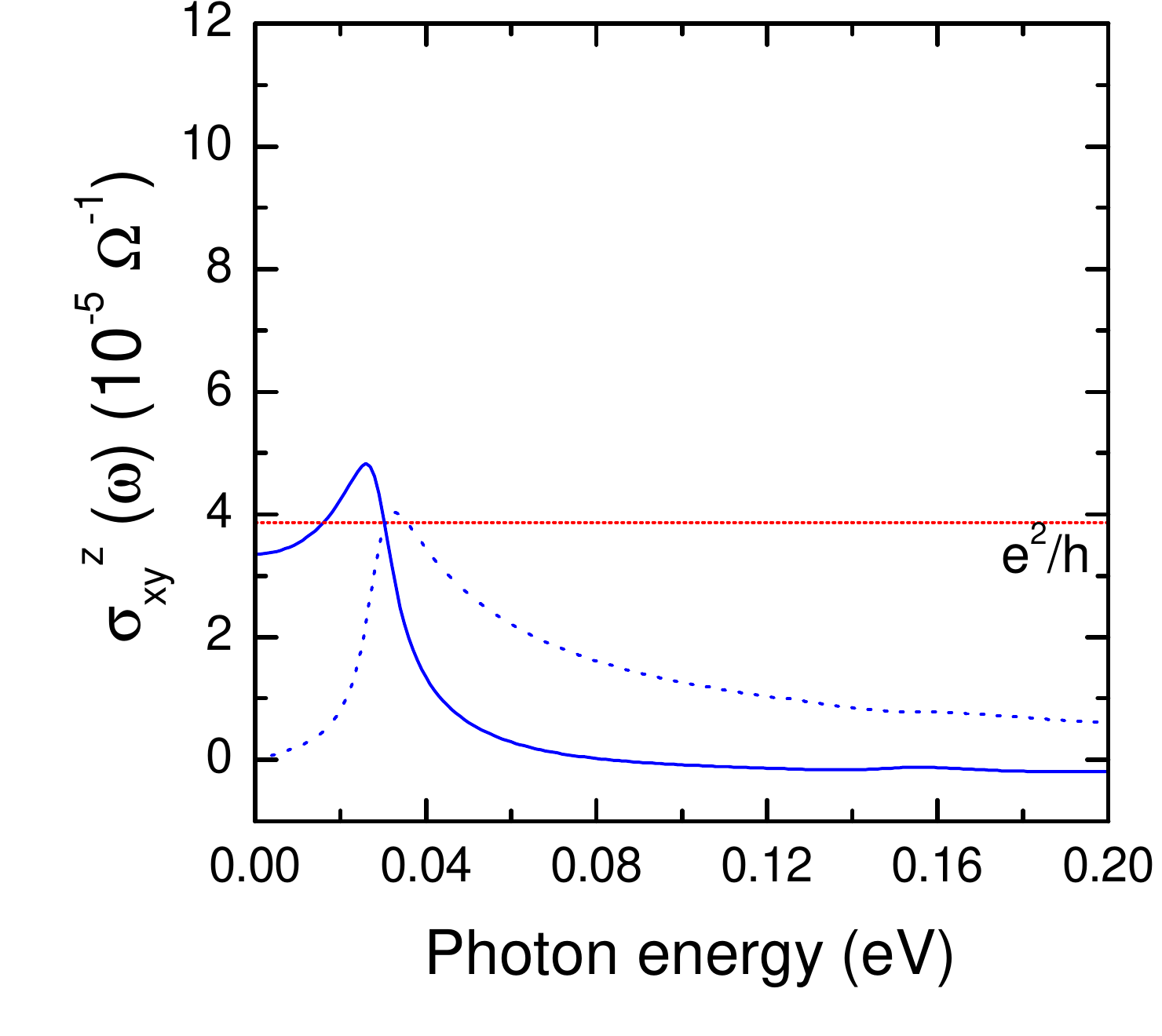}}
	\subfigure[{} 1S-WS$_2$]{\includegraphics[scale=0.28]{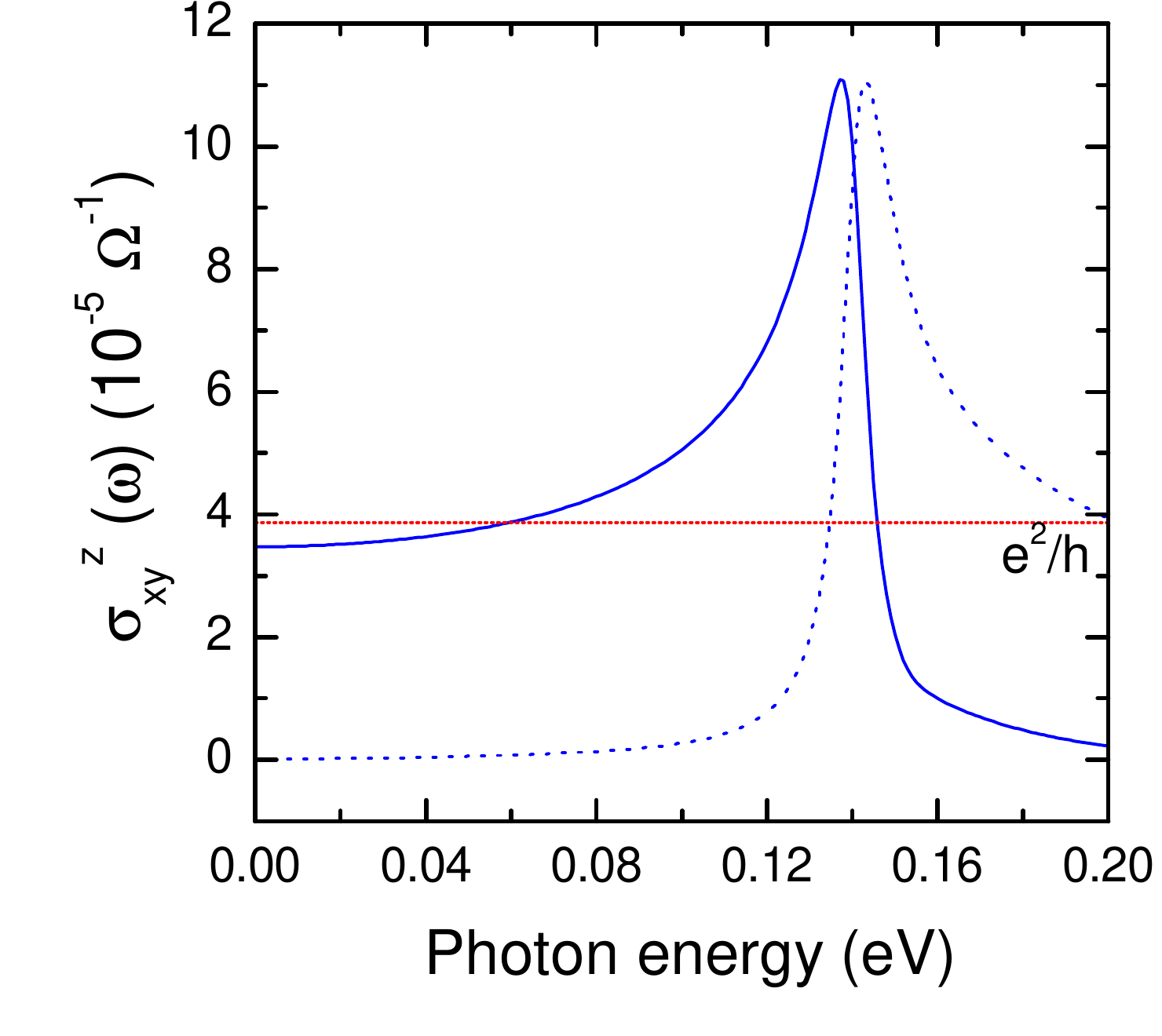}}
	\subfigure[{} 1T'-MoS$_2$]{\includegraphics[scale=0.28]{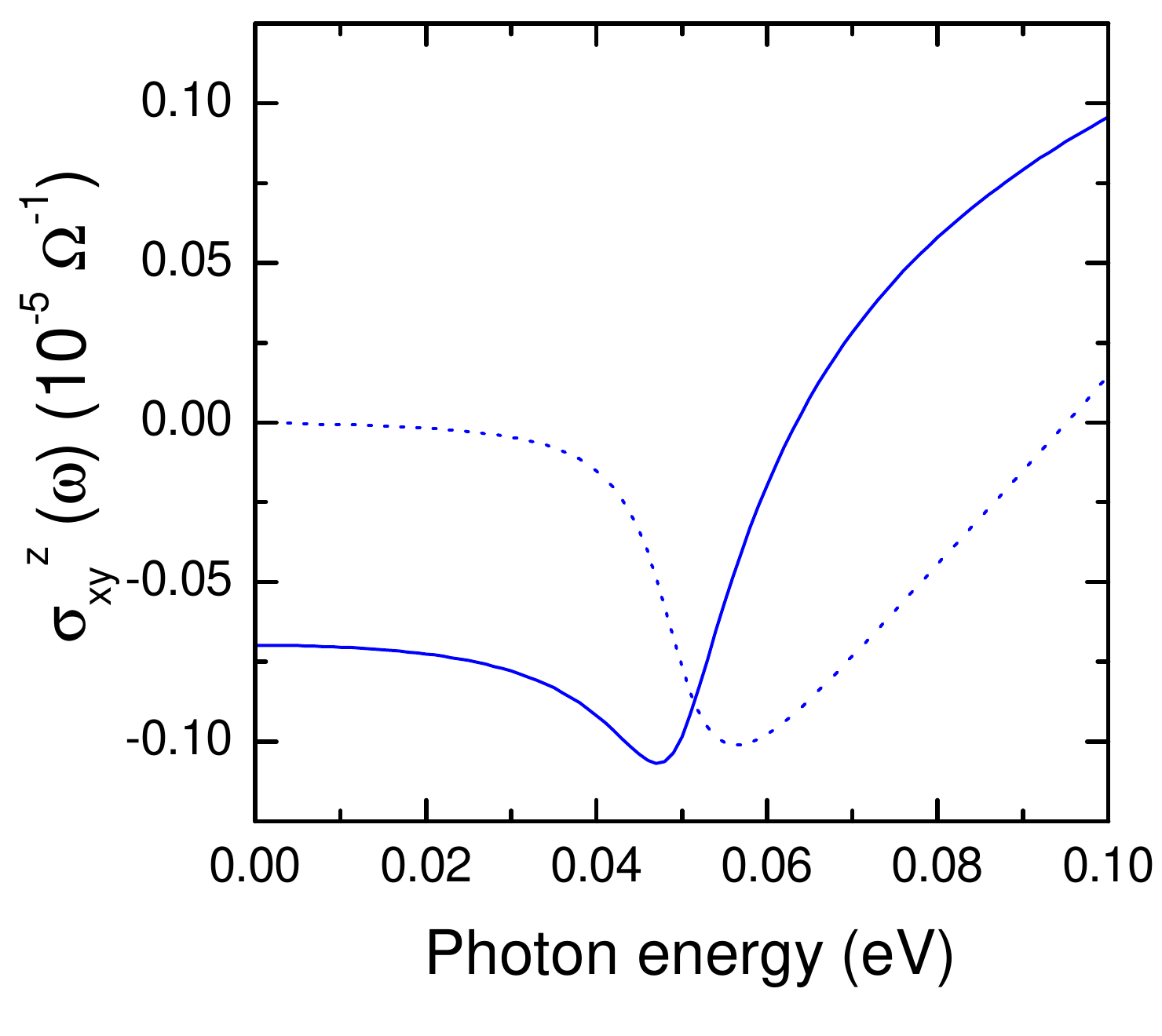}}
	\subfigure[{} 1T'-WS$_2$]{\includegraphics[scale=0.28]{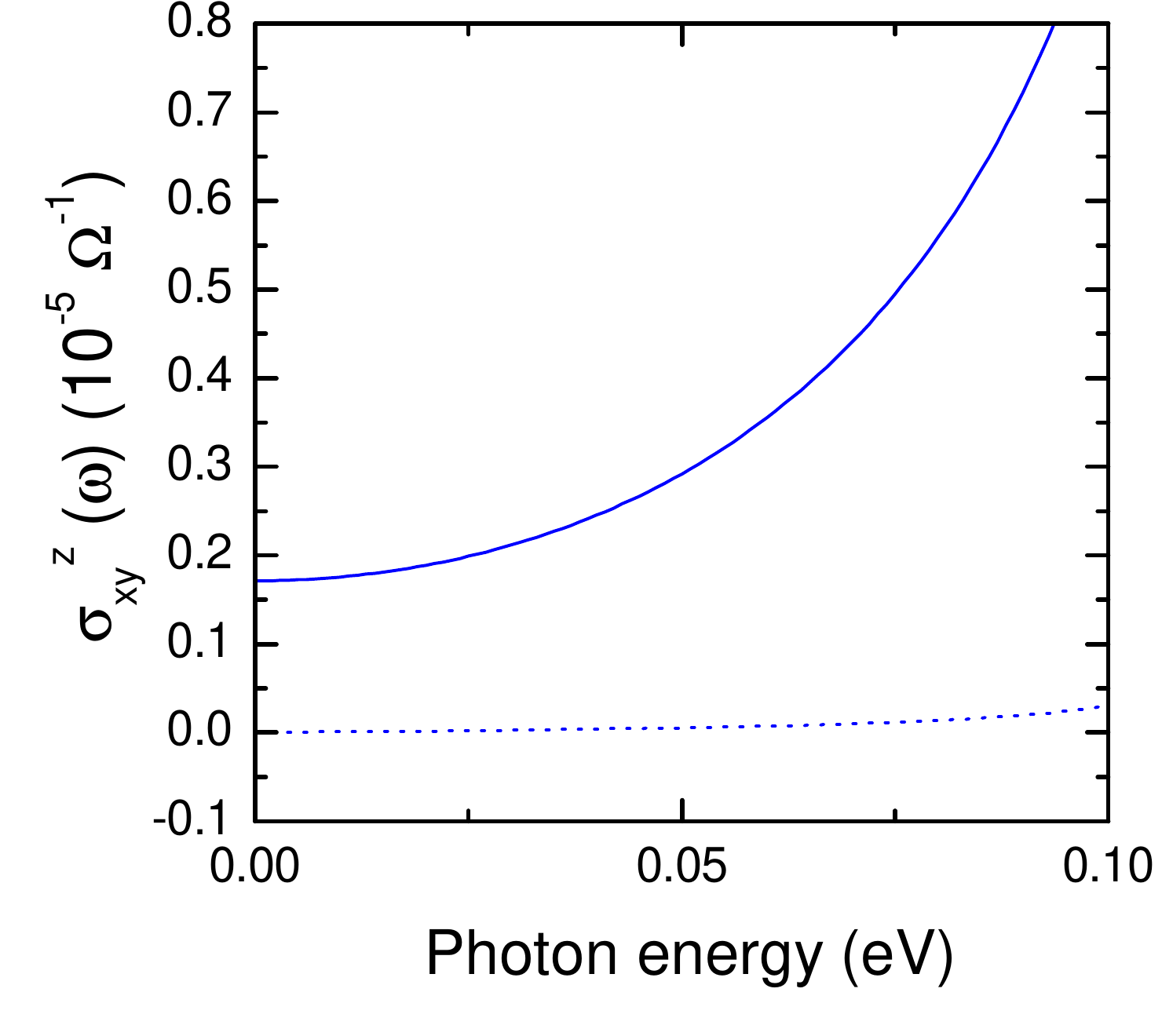}}
	\caption{\label{sigma-freq2} Spin Hall conductivity $\sigma^z_{xy}(\omega)$ (real part: solid lines, imaginary part: dashed lines) as a
		function of photon energy for (a) 1S-MoS$_2$, (b) 1S-WS$_2$, (c) 1T'-MoS$_2$ and (d) 1T'-WS$_2$. The sample temperature is set to 0 K. The broadening parameter is fixed at $\eta = 0.005$ eV.}
\end{figure}

\bibliography{/home/filipe/Dropbox/artigos/library}